\documentclass[12pt,review]{elsarticle}

\usepackage{lineno,hyperref}
\usepackage{bm}
\usepackage{amsfonts}
\usepackage{amssymb}
\usepackage{amsmath}
\usepackage[boxed]{algorithm2e}
\usepackage{isomath}

\usepackage{booktabs} 
\usepackage{siunitx}  

\usepackage{scalerel,stackengine}
\stackMath
\newcommand\reallywidehat[1]{%
\savestack{\tmpbox}{\stretchto{%
  \scaleto{%
    \scalerel*[\widthof{\ensuremath{#1}}]{\kern-.6pt\bigwedge\kern-.6pt}%
    {\rule[-\textheight/2]{1ex}{\textheight}}
  }{\textheight}%
}{0.5ex}}%
\stackon[1pt]{#1}{\tmpbox}%
}

\usepackage{stmaryrd} 
\usepackage[usenames]{color}
\usepackage{epsfig}
\usepackage{graphicx}
\usepackage{psfrag}

\graphicspath{{Figures/}}
\usepackage{float}
\floatstyle{boxed}
\newfloat{algorithm}{htb}{loa}
\floatname{algorithm}{Algorithm}
\usepackage{caption}
\usepackage{subcaption}
\usepackage{color}
\usepackage{xcolor}
\usepackage{multirow}

\modulolinenumbers[5]
\journal{Authors}
\usepackage{natbib}

\usepackage{subfloat}
\usepackage{multirow}
\usepackage{hyperref}
\usepackage{pict2e}
\usepackage{mdframed}
\newmdenv[
  skipabove=\topsep,
  skipbelow=\topsep
]{siderules}

\usepackage[left = 2cm, right=2.5cm, top=2cm, bottom =2cm]{geometry}

\begin{document}
\begin{frontmatter}




\title{An anisotropic cohesive fracture model: advantages and limitations of length-scale insensitive phase-field damage models} 



\author{Shahed Rezaei$^{1}$, Ali Harandi$^{2}$, Tim Brepols$^2$, Stefanie Reese$^2$} 
\address{$^1$Mechanics of Functional Materials Division, Institute of Materials Science,\\ Technische Universität Darmstadt, Darmstadt 64287, Germany}
\address{$^2$Institute of Applied Mechanics,\\ RWTH Aachen University, D-52074 Aachen, Germany}
\corref{s.rezaei@mfm.tu-darmstadt.de}

\begin{abstract}
The goal of the current work is to explore direction-dependent damage initiation and propagation within an arbitrary anisotropic solid. 
In particular, we aim at developing anisotropic cohesive phase-field (PF) damage models by extending the idea introduced in \cite{REZAEI2021a} for direction-dependent fracture energy and also anisotropic PF damage models based on structural tensors. 
The cohesive PF damage formulation used in the current contribution is motivated by the works of \cite{LORENTZ201120, wu2018, GEELEN2019}. The results of the latter models are shown to be insensitive with respect to the length scale parameter for the isotropic case. This is because they manage to formulate the fracture energy as a function of diffuse displacement jumps in the localized damaged zone. 
In the present paper, we discuss numerical examples and details on finite element implementations where the fracture energy, as well as the material strength, are introduced as an arbitrary function of the crack direction. Using the current formulation for anisotropic cohesive fracture, the obtained results are almost insensitive with respect to the length scale parameter. The latter is achieved by including the direction-dependent strength of the material in addition to its fracture energy.
Utilizing the current formulation, one can increase the mesh size which reduces the computational time significantly without any severe change in the predicted crack path and overall obtained load-displacement curves. We also argue that these models still lack to capture mode-dependent fracture properties. Open issues and possible remedies for future developments are finally discussed as well. \\
\end{abstract} 

\end{frontmatter}

\newpage
\section{Introduction}  
\begin{figure}[H]
    \centering
    \includegraphics[width=0.75\linewidth]{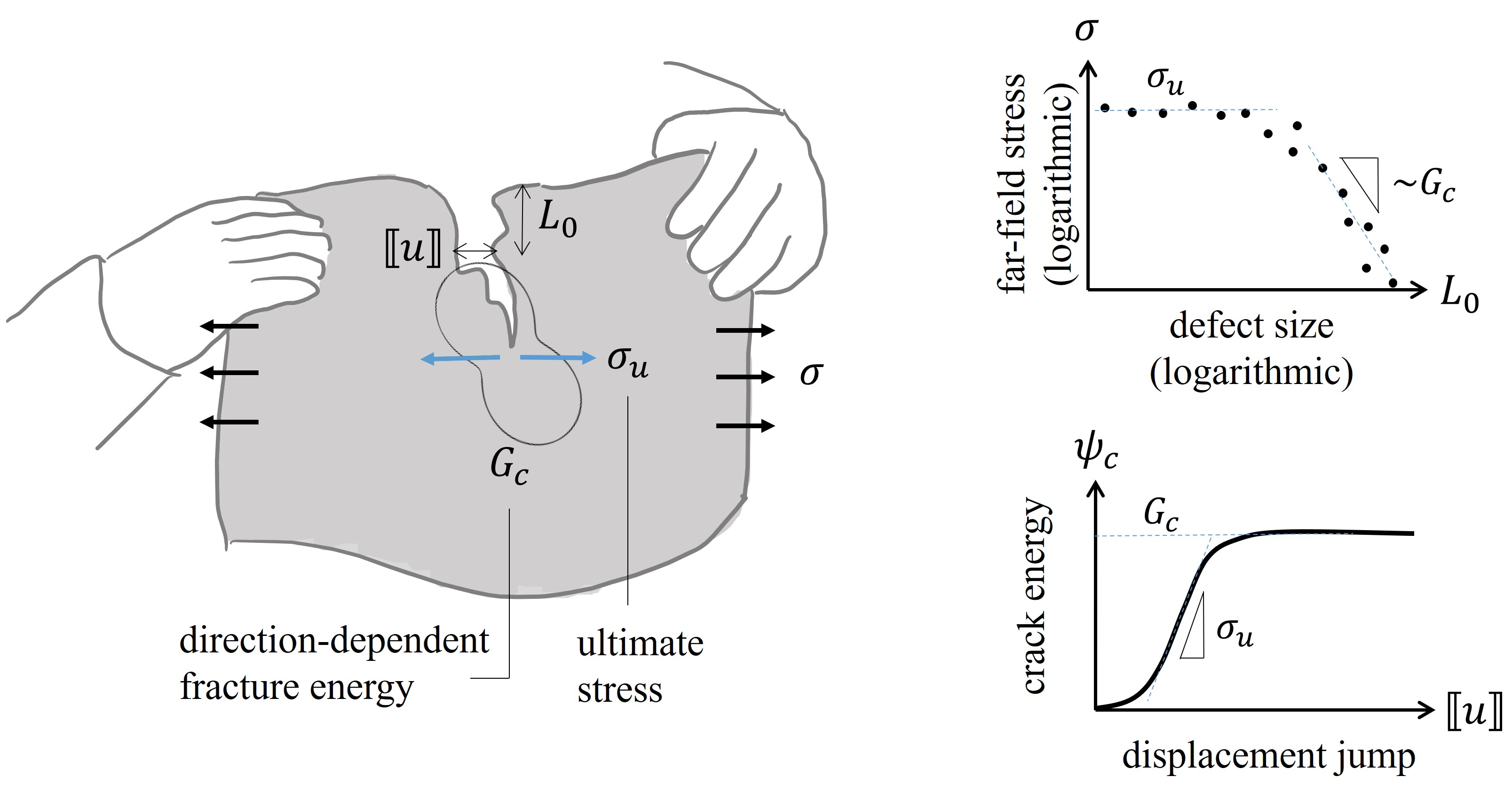}
    \caption{Left: crack direction is under the influence of the direction-dependent fracture energy with an arbitrary complex shape. Right: the ultimate stress of the material controls when the crack initiates  \cite{TenneBourdin,Verhossel2013}}.
    \label{fig:intro}
\end{figure}

Understanding and modeling damage is one among many challenging aspects in computational mechanics which has led to a significant amount of research in the recent decades. One can summarize the main questions into (1) when do cracks start to grow and (2) in which direction or where in the solid do they tend to propagate (see Fig.~\ref{fig:intro}). In the early work of Griffith \cite{Griffith1921}, an energy criterion for crack propagation is mentioned which is well accepted for predicting brittle fracture in materials/structures with an initial crack or defects. An alternative method is proposed by Barenblatt \cite{barenblatt1962mathematical} who introduced the the concept of cohesive fracture at the crack tip. Here, in addition to fracture energy, the maximum strength of the material is treated as a material property and used for predicting the crack nucleation. 

Being able to differentiate between several phases through a smooth transition, the phase-field (PF) damage model has shown a great potential to address damage in solids. This is achieved by introducing an order parameter to describe the transition from intact material to the fully damaged one \cite{FRANCFORT1998, Bourdin2000, Miehe2010a}. The phase-field damage model has proven to be an elegant tool, especially when multi-phases of a material \cite{SCHNEIDER2016186}, or a multiphysics problem \cite{Xu2016, Paneda2018, MOSHKELGOSHA2021107403} are considered. For an overview of the model and a survey of recent advances, see \cite{BUI2021107705}.

Despite the interesting features of the PF damage model, one needs to treat the internal length scale parameter with care. The length scale parameter controls the width of fracture process zone and 
its value is usually small with respect to the structure's size \cite{Steinke2017}. Utilizing the PF damage model, same amount of energy is dissipated upon crack progress, independent of the internal length scale parameter (see also \cite{Linse2017} for further studies). Nevertheless, it was also shown that this parameter directly influences the overall response of the structure (e.g. measured force-displacement). Therefore, in the standard PF damage model, the length scale parameter can not be seen as a pure numerical parameter \cite{AMOR2009, KUHN2014051008, BORDEN201277},  
Utilizing a simplified analytical solution, one can show the relationship between the length scale parameter and the strength of the material \cite{Nguyen2016, Zhang2017}. In other words, by utilizing a standard PF damage formulation, we are restricted in choosing the length-scale parameter. This could be problematic when it comes to simulations on a small scale as the relatively wide damage zone may create some boundary effects. Note that since the value of the length scale parameter is linked to the material properties, one is not to allowed to choose smaller values. Next, we look at the cohesive nature of fracture. 

In addition to the fracture energy value, information on how fracture energy reaches its peak value is essential. Taking the latter point into account, one is able to improve the standard PF or even continuum damage models \cite{LORENTZ201120, Verhossel2013}.
Interestingly enough, such dependency is investigated already in the context of cohesive zone (CZ) modeling \cite{Mergheim2005, Rezaei2017}. The constitutive relation for the CZ model is defined by employing a so-called traction separation (TS) relation. It was shown that CZ models can be calibrated based on information from lower scales down to atomistic level \cite{Rezaei2019, Rezaei2020}. 



The above arguments are summarized in Fig.~\ref{fig:intro}. On the left-hand side, a single notched specimen is shown. The questions that we would like to address are when the crack starts to grow and where in the solid it propagates (in which direction). It is shown that the fracture energy is in general direction-dependent and there are in general certain preferential directions for the crack propagation \cite{REZAEI2021a}. The latter point mainly determines the crack propagating direction. It is also known that when the length of the initial defect $L_0$ is large enough (compared to the specimen size), the dominant factor in crack propagation is the fracture energy value $G_c$ (or fracture toughness \cite{Gibson_Shahed}). On the other hand, when the initial crack length vanishes, the strength of the material (also known as ultimate tensile stress), is the parameter that controls the crack initiation. 
 
Interestingly enough, it was shown that the PF damage model can capture such a nonlinear transition \cite{KUMAR2020104027, MOLNAR2020102736}. On the other hand, the question remains on how can we treat the internal length scale in this situation to make the formulation independent of it. The latter point will be examined in what follows. Finally, the idea of this work is to combine the above-mentioned features in one formulation to address anisotropic cohesive fracture in solids.

\subsection{Phase-field modeling of cohesive fracture}
By integrating the cohesive response of fracture in the PF damage model, it was shown that one can omit the direct influence of the internal length scale parameter on the overall results. Instead, one should introduce the maximum strength as an additional model parameter.

Available works on cohesive phase-field fracture are divided into two mainstreams. The first category can be seen as an extension of the classical CZ model where instead of a sharp or interphase interface one deals with a diffuse damage zone. Note that similar to CZ models, the interphase position in these works is known in advance. Verhoosel and de Borst~\cite{Verhossel2013} included the idea behind cohesive fracture in PF damage models by introducing an extra auxiliary field for the displacement jump. See also \cite{NGUYEN2016567, Tarafder2020}, where the authors described the sharp interface by employing a diffuse zone which is the idea behind phase-field theory. 

In the second category, the focus is on modifying standard PF damage models in a way that they can represent the cohesive nature of fracture. Note that one is still able to predict an arbitrary crack path using such methods. Motivated by~\cite{LORENTZ20111927, LORENTZ201120}, new forms of energetic formulations and functions were developed through which the cohesive fracture properties are taken into account. These functions were recently used by \cite{Tupeck2016, wu2017, GEELEN2019} in PF damage models. Wu and Nguyen \cite{wu2018} presented a length scale insensitive PF damage model for brittle fracture. Utilizing a set of characteristic functions, the authors managed to incorporate both the failure strength and the traction-separation relation, independent of the length scale parameter. Geelen et al. \cite{GEELEN2019} extended the PF damage formulation for cohesive fracture by making use of a non-polynomial degradation function. The interested reader is referred to \cite{Fang2020, FREDDI2017156} for more details.

\subsection{Anisotropic fracture: direction-dependent fracture energy}
Various microstructural features such as grain morphology or fiber orientation have a huge impact on the material's fracture properties. Such material features influence the crack direction within an arbitrary solid. Anisotropic crack propagation can be linked to a direction-dependent fracture energy function \cite{Hakim2005, REZAEI2021a}. Note that anisotropic elasticity is not enough to fully capture anisotropic damage behavior \cite{Hakim2005, GAO2017330}. 

Anisotropic crack propagation in the context of phase-field models is based on two mainstreams. The first approach focuses on introducing structural tensors into the formulation which act on the gradient of the damage variable, forcing the crack to propagate in certain directions \cite{Teichtmeister17, LI2019502}. This approach in consideration with only one scalar damage variable cannot model arbitrary anisotropic fracture. Utilizing a second-order structural tensor will result in a fracture energy distribution which only has one major preferential direction for the crack. Therefore, this method is also known to cover only \textit{weak anisotropy}. By utilizing higher-order damage gradient terms and a fourth-order structural tensor, one can simulate the so-called \textit{strong anisotropy} with two preferential crack directions. The latter point is important in crystals with cubic symmetry \cite{Li2015, KAKOURIS2019112503, Teichtmeister17}. Beside being computationally more demanding, the latter approaches are limited to certain shapes (distribution) of the fracture energy function. To take into account more complicated fracture energy patterns, a promising extension would be to make use of several damage variables. Nguyen et al. \cite{NGUYEN2017279} introduced multiple PF damage variables. Each damage variable is responsible for stiffness degradation in a certain direction (see also \cite{NguyenYv2019}). Nevertheless, this approach also increases the computational cost and opens up other questions, e.g. on how different damage variables should influence the initial material's elastic stiffness.

Interestingly enough, the (fracture) surface energy of a crystalline solid might become a non-trivial function of orientation \cite{EGGLESTON200191, Zhang2017bat, BRACH201996, REZAEI2021a}. 
Hossain et al. \cite{HOSSAIN2018118} presented the influence of crystallographic orientation on toughness and strength in graphene. The latter observations suggest that in general, one has to deal with an arbitrary complex distribution for the fracture toughness of the material. Therefore, in the second strategy, the fracture energy parameter may be defined as a function of the crack direction \cite{REZAEI2021a}. Very recently, the idea behind cohesive fracture is also combined with anisotropic crack propagation using the PF damage model \cite{pillai2020, ZHANG2020113353, MANDAL2020105941}. Still, fundamental studies are required on why the length scale insensitive PF damage model might be necessary. 



The outline of the current contribution is as follows. In section 2, the formulation for the anisotropic insensitive phase-field damage model is discussed. In section 3, the discretization of the problem for implementation in the finite element method is covered.  Numerical examples are then presented in section 4. Finally, conclusions and an outlook are provided.

\section{Anisotropic phase-field model for cohesive fracture}
\begin{figure}[H]
    \centering
    \includegraphics[width=0.9\linewidth]{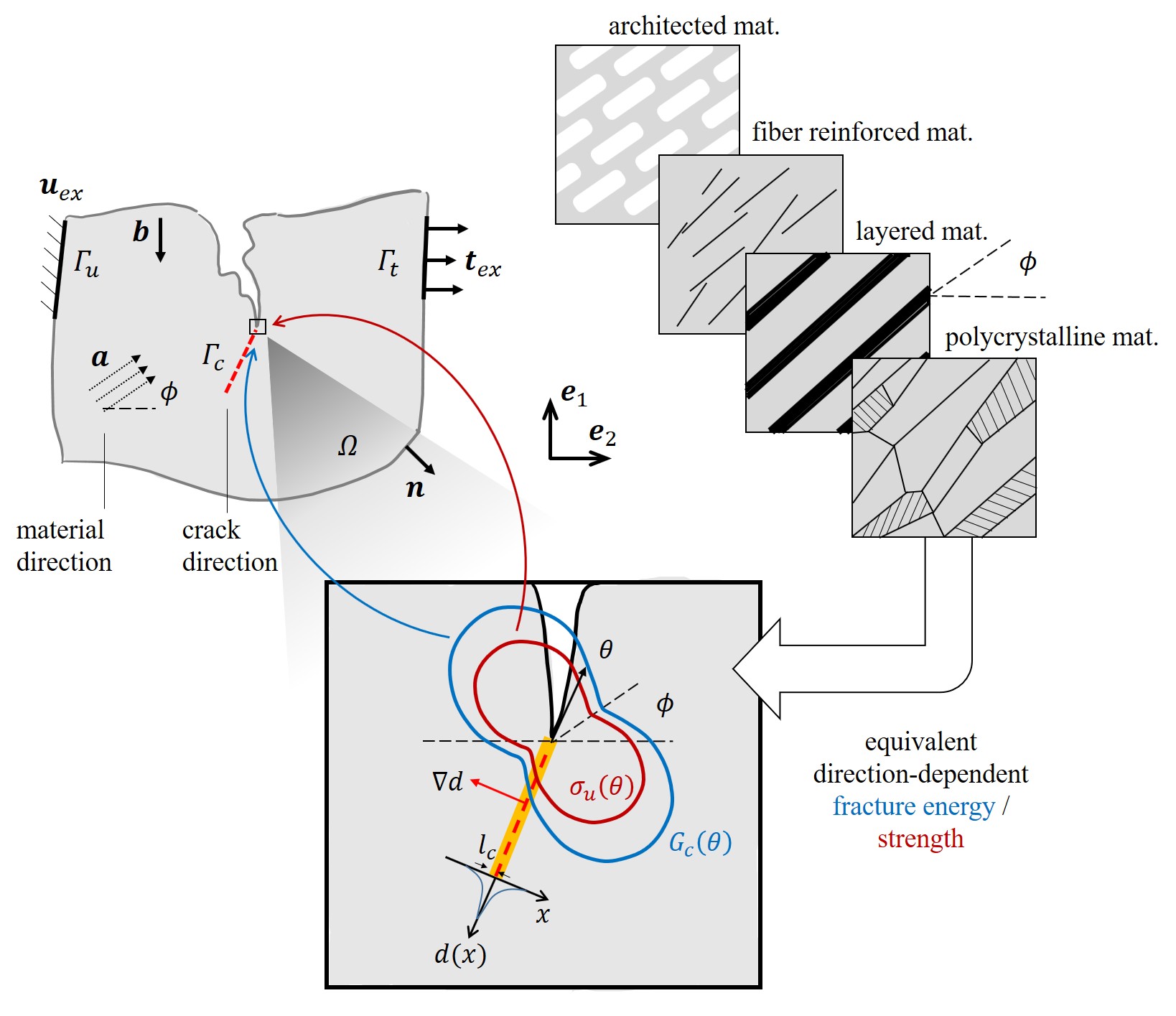}
    \caption{Configuration of a general elastic body and different applied boundary conditions.}
    \label{fig:body}
\end{figure}

In the left part of Fig.~\ref{fig:body}, the configuration of an anisotropic elastic body $\Omega$ is shown.The specific material direction $\phi$ (e.g. fibers' direction or grains' orientation) is represented by the vector $\bm{a}$.  
According to the right hand side of Fig.~\ref{fig:body}, the direction-dependent fracture energy, strength and elasticity of the material can be traced back to its material microstructure. The main idea behind the current formulation is to take such properties into account in the PF damage formulation. The position and displacement vector of an arbitrary point are represented by $\boldsymbol{x}$ and $\boldsymbol{u}$, respectively. The applied displacement, traction and body force vectors are denoted by $\bm{u}_{ex}$, $\boldsymbol{t}_{ex}$ and $\boldsymbol{b}$, respectively.  

\begin{figure}[H]
    \centering
    \includegraphics[width=1.0\linewidth]{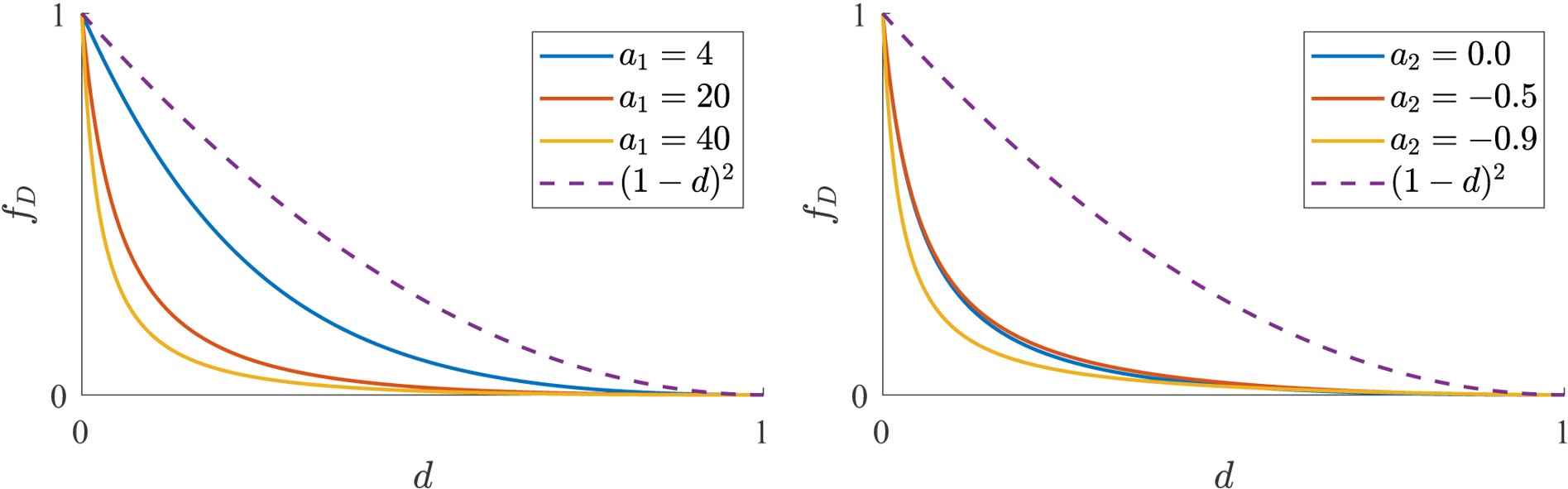}
    \caption{Influence of parameters $a_1$ and $a_2$ on the degradation functions (Eq.~\ref{f_D}). In the left and right hand side we have $a_2=-0.9$ and $a_1=40$, respectively}
    \label{fig:fd}
\end{figure}

The sharp crack $\Gamma_c$ is represented by a diffuse damage field $d(x)$. The width of the damage zone is controlled by the length scale parameter $l_c$.
The internal energy density of the system $\psi$ is divided into an elastic part $\psi_e$ and a damage part $\psi_c$. The latter shows the additional energy of the newly created surfaces upon cracking: 
\begin{equation}
  \label{internalenergy}
\psi(\bm{\epsilon},d,\nabla d)=\psi_e (\bm{\epsilon},d) + \psi_c(d,\nabla d),
\end{equation}
where $\boldsymbol{\epsilon}=0.5~(\nabla \boldsymbol{u} + \nabla \boldsymbol{u}^T)$ is the strain tensor for small deformations.
From Eq.~\ref{internalenergy}, it becomes clear that in the PF damage formulation, one deals with two separate fields, namely, the displacement vector $\bm{u}$ and the damage parameter $d$. These two independent variables are strongly coupled together. The elastic energy part takes the standard form
\begin{align}
\label{elasticenergy}
\psi_e = \dfrac{1}{2} \bm{\epsilon}:\mathbb{C}:\bm{\epsilon}.
\end{align}
The fourth order elastic stiffness tensor $\mathbb{C}$ is influenced by damage according to the following split (see \cite{AMOR2009}):
\begin{align}
\label{Cfourth}
\mathbb{C} &= f_D~\mathbb{C}_0 + (1-f_D)~\mathbb{P},\\
\mathbb{C}_0 &= \lambda \bm{I}\otimes\bm{I} + 2 \mu \mathbb{I}^s,\\
\label{amormethod}
\mathbb{P} &= k_0~\text{sgn}^{-}(\text{tr}(\bm{\epsilon}))~ \bm{I} \otimes \bm{I}.
\end{align}
Here, $(\mathbb{I}^s)_{ijkl}=\dfrac{1}{2}(\delta_{ik}\delta_{jl}+\delta_{il}\delta_{jk})$ is the symmetric fourth-order identity tensor. The second order identity tensor is defined as $(\bm{I})_{ij}=\delta_{ij}$.
Considering Young's Modulus $E$ and the Poisson ratio $\nu$ for elastic isotropic materials, $\lambda=\dfrac{E\nu}{(1+\nu)(1-2\nu)}$ and $\mu=G=\dfrac{E}{2(1+\nu)}$ are the Lame constants. In Eq.~\ref{Cfourth}, the introduced damage function $f_D$ degrades the initial (undamaged) material stiffness $\mathbb{C}_0$. The degradation function $f_D$ plays a significant role in the cohesive behavior. According to \cite{wu2018, LORENTZ201120}, for bilinear cohesive laws this function takes the form:
\begin{equation}
  \label{f_D}
f_D = \dfrac{(1-d)^2}{(1-d)^2+a_1d\,+\,a_1\,a_2\,d^2}.
\end{equation}
In the above equation, $a_1$ and $a_2$ are constant model parameters that have to be chosen. They are determined considering the cohesive properties of the model, e.g. the ultimate stress before damage initiation and the value for strain at the fully broken state (see Appendix A and \cite{wu2018}). Note that there are certainly other choices for the damage function as well (see for example \cite{GEELEN2019}). In general, the damage function takes the value one and when damage approaches one (crack is fully developed), this function vanishes. The degradation function in Eq.~\ref{f_D} is plotted in Fig.~\ref{fig:fd} and compared against some classical choices.

Furthermore, we require a split in tensile and compressive elastic energy parts to avoid material damage under compressive loading. Here, the approach of \cite{AMOR2009} is considered, where we take into account only the positive volumetric part of the strain to damage the material. The sign function $\text{sgn}^{-}(\bullet)=(\bullet - | \bullet |)/2$, only takes the negative part of its argument. The fourth-order projection tensor $\mathbb{P}$ is defined in Eq.~\ref{amormethod} to exclude material parts in compression. Moreover, the bulk modulus of the material is defined as $k_0~= \lambda+\dfrac{2}{3}~\mu$. This approach is simple to implement and yet effective in many applications especially when it comes to initially anisotropic materials. There are more advanced splits available in the literature. 

The energy for creating a new pair of surfaces per unit length is described as fracture energy $G_c$. Therefore, in the PF damage formulation, we have \cite{MIEHE201027,AMOR2009}
\begin{equation}
\label{psiC}
    \int_{\Omega} \psi_c(d,\nabla d)~dV = 
    \int_{\Gamma_c}G_c~dA
\end{equation}
In this work, we take this concept further and make this energy dependent on the direction of the crack, i.e. $G_c(\theta)=g_c(\nabla d)$ \cite{REZAEI2021a}. Here it is assumed that the crack direction is perpendicular to the damage gradient $\nabla d$ (see Fig. \ref{fig:body}).
Note that in the PF damage model, the damage gradient vector can not easily be defined when there has no damage evolved in the system yet. As will be discussed later, for a better convergence in the finite element calculations, we will apply some numerical treatments.
According to \cite{REZAEI2021a}, the angle $\theta$ which represents the crack direction is defined according to
\begin{equation}
  \label{theta}
\theta = \text{atan}\left(\dfrac{\nabla d \cdot \bm{e}_2}{\nabla d \cdot \bm{e}_1}\right)-\dfrac{\pi}{2}.
\end{equation}

In what follows, we also review the formulation for anisotropic PF damage model using structural tensors. Note that for the latter approach one needs a constant value for the fracture energy (i.e. $G_{c,0}$). The energy required for creating a crack $\Gamma_c$ is regularized over the volume such that we write \cite{FRANCFORT1998, MIEHE201027}:
\begin{equation}
  \label{regularizedcrackenergy}
\psi_c(d,\nabla d) = \begin{cases}
\psi_{c,s} = G_{c,0}~\gamma_s(d,\nabla d),~~~~~\text{for structural anisotropy} \\
\psi_{c,a} = G_c(\theta)~\gamma_a(d,\nabla d),~~~\text{for arbitrary anisotropy}
\end{cases},
\end{equation} 
where $\gamma_s$ and $\gamma_a$ are the crack density function for the case of structural anisotropy and arbitrary anisotropy, respectively. A more detailed definition is given in the following part. In Eq.~\ref{regularizedcrackenergy}, the direction-dependent fracture energy is represented by $G_c(\theta)$.

\textbf{Remark 1.} The statement that the vector $\nabla d$ is orthogonal to the crack plane is an approximation and does not hold in small regions at the crack tip. However, this effect is quite localized. Although the influence of the latter point might be negligible but further studies on this point would be interesting. 


\subsection{Modeling anisotropic fracture with structural tensors}
To model anisotropic fracture, it is common in the literature to use a second-order structural tensor. The crack density function in this particular case is written as:
\begin{equation}
  \label{stcrackdensity}
\gamma_s(d,\nabla d)= \dfrac{1}{c_0\, l_c}~\omega(d)~+~\dfrac{l_c}{c_0}~\nabla d \cdot \bm{A} \cdot \nabla d.
\end{equation}
In Eq.~\ref{stcrackdensity}, the scalar parameter $l_c$ is the internal length scale and represents the width of the localized (damage) zone. Furthermore, the crack topology function is represented by $\omega(d)$. The scalar parameter $c_0=4\int_0^1\sqrt{\omega(d)}\text{d} s$ is obtained so the integration of the crack energy over volume represent the material fracture energy $G_c$ (see Eq.~\ref{psiC} and Appendix B). Similar to the damage function $f_D$, There are several choices for the crack topology function. For the cohesive PF damage model, we focus on the following form, through which we will have $c_0=\pi$ \cite{wu2018}.
\begin{equation}
  \label{topologyfunc}
\omega(d)= 2d - d^2.
\end{equation}



The second order structural tensor $\bm{A} = \boldsymbol{I}+\alpha~\boldsymbol{a}\otimes\boldsymbol{a}$ which is constructed based on the vector $\boldsymbol{a}$, penalizes the crack direction at a certain angle \cite{Teichtmeister17}. This angle is in accordance with the direction of the vector $\boldsymbol{a}=[\cos(\phi)~\sin(\phi)]^T$. Therefore, one can write
\begin{equation}
  \label{Atensor}
\bm{A} = \boldsymbol{I}+\alpha~\begin{bmatrix}
\cos(\phi) \\ \sin(\phi) 
\end{bmatrix} \begin{bmatrix}
\cos(\phi) & \sin(\phi) 
\end{bmatrix} =  \boldsymbol{I}+ \alpha \begin{bmatrix}
 \cos^2(\phi) & \cos(\alpha) \sin(\phi) \\ \cos(\phi) \sin(\phi)& \sin^2(\phi) 
\end{bmatrix}.
\end{equation}
In the above equation, the scalar parameter $\alpha$ determines the contribution of the preferential directions in the energy term. In other words, the higher the parameter $\alpha$ is, the more energy we require to form a crack perpendicular to the direction pointed by vector $\boldsymbol{a}$. Moreover, the angle $\phi$ denotes the preferred direction (e.g. grains, fibers and etc.). Since in this work we focus on geometrically linear setting, the angle $\phi$ is kept constant through out the derivation and further calculations.

After some simplifications (see Appendix C), one can obtain the following relation for the anisotropic fracture surface energy utilizing the second-order structural tensor introduced in Eq.~\ref{Atensor}:
\begin{equation}
\label{anisostpsic}
\psi_{c,s} = G_{c,0}~\gamma_s = \dfrac{G_{c,0}}{\pi l_c}~\omega(d) + \dfrac{G_{c,0}~l_c}{\pi} \, ||\nabla d||^2 \left(1+\alpha \sin^2 (\theta - \phi )\right).
\end{equation} 
Note that the angle $\theta = \text{atan}\left(\dfrac{\nabla d \cdot \bm{e}_2}{\nabla d \cdot \bm{e}_1}\right)-\dfrac{\pi}{2},$ is the crack direction and given in dependence of $\nabla d$ (see Eq.~\ref{theta}). The second term of crack surface energy in Eq.~\ref{anisostpsic} is the response term for the directional dependent fracture energy. 
Parameters $G_{c,0}$, $\alpha$ and $\phi$ are model input parameters. This formulation is also known as a weak anisotropy \cite{Teichtmeister17}. Although in this work we will focus on this particular formulation, later on, we will introduce the formulation with arbitrary anisotropy as well.

\textbf{Remark 2.} Enhancing the crack density function with second-order structural tensors showed a great performance in simulating anisotropic crack propagation in various applications. However, by considering only one damage variable, such an extension is not general enough for materials with strong anisotropy. Utilizing higher-order terms in the crack density function such as $\gamma = \dfrac{1}{2l_c}d^2 + \dfrac{l_c}{4}{\nabla d} \cdot {\nabla d} + \dfrac{l_c^3}{32} {\nabla}^2{d}:\mathbb{A}:{\nabla}^2{d}$ is an interesting option. In the latter formula, $\mathbb{A}$ is a fourth-order structural tensor which is defined employing preferable directions for the crack (see \cite{Teichtmeister17}). These extensions can be even combined with several damage variables to take into account more complex anisotropic behavior \cite{nguyen2017a}. In what follows, we keep the crack density function $\gamma$ to be the same as a standard one and the amount of fracture energy is directly plugged in through the function $g_c$. This is motivated based on the arbitrariness of the fracture energy for a solid (see \cite{REZAEI2021a}).

For the further derivation of the model, the following thermodynamic forces are introduced. First, according to Eq.~\ref{elasticenergy} and Eq.~\ref{Cfourth}, the stress tensor as a conjugate force to the strain tensor reads:
\begin{equation}
  \label{stressconjugate}
\dfrac{\partial \psi_e}{\partial \bm{\epsilon}}=\bm{\sigma}=\mathbb{C}:\bm{\epsilon}=f_D~\mathbb{C}_0 :\bm{\epsilon} + (1-f_D)~\mathbb{P}:\bm{\epsilon}.
\end{equation}
Furthermore, the damage driving force $Y$ from elastic energy reads:
\begin{equation}
  \label{damagedrivingforce}
\dfrac{\partial \psi_e}{\partial d}=-Y=\dfrac{\text{d} f_D}{\text{d} d}~\dfrac{1}{2}~\bm{\epsilon}:\mathbb{C}_h:\bm{\epsilon},
\end{equation}
where $\mathbb{C}_h=\mathbb{C}_0-\mathbb{P}$. By using the Euler-Lagrange procedure, the variational derivative of the total energy with respect to the displacement field results in the standard mechanical equilibrium \cite{Miehe2010a, Teichtmeister17, SCHNEIDER2016186}:
\begin{equation}
\label{deltau}
\delta_{\bm{u}}\psi = \partial_{\bm{u}}\psi - \text{div}(\partial_{\nabla {\bm{u}}} \psi) = 0 \Rightarrow \begin{cases}
\text{div}(\bm{\sigma)} + \bm{b}=  \bm{0}~~\text{in}~\Omega \\
\bm{\sigma} \cdot \bm{n}=\bm{t}_{ex}~~~~~~~~\text{on}~\Gamma_t \\
\bm{u}=\bm{u}_{ex}~~~~~~~~~~~~\text{on}~\Gamma_{\bm{u}}
\end{cases}.
\end{equation}
Next, the variational derivative with respect to the damage field is considered \cite{Teichtmeister17, SCHNEIDER2016186}.
\begin{equation}
  \label{deltadsttensor}
\delta_d \psi = \partial_d \psi - \text{div}(\partial_{\nabla d} \psi) = 0 \Rightarrow \begin{cases}
\dfrac{G_{c,0}}{l_c \pi} \omega^{\prime}-\text{div}\left(\dfrac{l_c G_{c,0}}{c_0} ~\bm{A}:\nabla d\right)-Y_{m,s}=0 ~\text{in}~\Omega \\
\nabla d \cdot \bm{n}=0~\text{on}~\Gamma_c
\end{cases}
\end{equation}
In above equations, we utilize the maximum damage driving force $Y_{m,s}$ to consider for damage irreversibility upon unloading. The expression for $Y_{m,s}$ is defined as maximum value between the undamaged elastic strain through the simulation time ($\psi_e^0(t)$) and the damage energy threshold ($\psi_{th}$) \cite{wu2018, MANDAL2020105941}:
\begin{equation}
  \label{Y_m}
-Y_{m,s} = f_D^{\prime}H_s = f_D^{\prime} \max_t (\psi_e^0(t),\psi_{th,s}).
\end{equation}
Here, the maximum value of stored undamaged elastic energy $\psi^0_e = \dfrac{1}{2} \bm{\epsilon}:\mathbb{C}_h:\bm{\epsilon}$ during the simulation time is denoted by $H_s$. The scalar parameter $H_s=\max_t (\psi_e^0(t),\psi_{th,s})$ is treated as a history variable throughout the simulation (see also \cite{Miehe2010a, Borst2015may, MIEHE2015449, wu2018}). 

The energy threshold $\psi_{th,s}$ ensures that damage remains zero as long as the elastic energy of the system is below this threshold. This is achieved based on the linear damage term in the damage topology function $\omega(d)$ (Eq.~\ref{topologyfunc}). See also section 2.3 and \cite{MIEHE2015449, wu2018}. The system's elastic energy right before onset of failure can be written in terms of the failure initiation strain $\varepsilon_0$ or the failure stress $\sigma_0$:
\begin{equation}
\psi_{th,s} = \dfrac{1}{2}E~\varepsilon^2_0 = \dfrac{1}{2E}\sigma_0^2.
\end{equation}

More explanations for the chosen relations in the above equation are provided at the end of this section. For the derivative of the damage function $f_D$ with respect to to the damage variable, we have
\begin{equation}
  \label{fdprime}  
f_D^{\prime}=\dfrac{\text{d} f_D}{\text{d} d} = \dfrac{-2 (1-d)\,(a_1d\,+\,a_1\,a_2\,d^2)-(1-d)^2\,(a_1\,+\,2a_1\,a_2\,d)}{\left((1-d)^2+a_1d+a_1a_2d^2\right)^2}.
\end{equation}
See also Eq.~\ref{f_D} and explanations provided afterwards for parameters $a_1$ and $a_2$. Based on the studies of \cite{wu2017, wu2018}, to represent a softening behavior similar to the bi-linear cohesive zone model, we choose the following form for these constant: 
\begin{align}
\label{a1_st_tensor}
a_{1,s} = \dfrac{4 E\,G_c}{\pi l_c ~\sigma_{0}^2},~~~~~
a_{2,s} = -0.5.
\end{align}

\textbf{Remark 3.} The scalar parameter $a_1$ in Eqs.~\ref{f_D}, \ref{a1_st_tensor} and \ref{a1_araniso} is defined to be length-scale dependent. As we will show later, this is one main reason why we have length-scale insensitive results for our cohesive phase-field damage model. In other words, via such a formulation one can control the maximum strength of the new material property (input) $\sigma_{u}=\sigma_0$.

\subsection{An arbitrary anisotropic fracture energy}
For this formulation, the crack density function $\gamma_a$, takes the standard form
\begin{equation}
  \label{arcrackdensity}
\gamma_a(d,\nabla d)= \dfrac{1}{c_0\, l_c}~\omega(d)~+~\dfrac{l_c}{c_0}~\nabla d \cdot \nabla d.
\end{equation}
Similar descriptions as for the previous case hold here for the parameter $c_0=\pi$, the length scale parameter $l_c$ as well as the crack topology function $\omega(d)$. Based on the recent work of the authors presented in \cite{REZAEI2021a}, to model anisotropic crack propagation, one can directly apply an arbitrary shape for the fracture energy function. Considering the crack angle $\theta$ (Eq.~\ref{theta}), it is suggested that the direction-dependent fracture energy function $G_c(\theta)$ can be obtained by summation over the frequency energy function. Here the sub-index $m$ which belongs to natural numbers represents the frequency number:
\begin{align}
\label{gcem}
G_c(\theta) &= \sum_m \kappa_m\left(1+\alpha_m~\sin^2\left(m(\theta+\theta'_m \right))\right), ~m\in \mathbf{N}.
\end{align}
The angle $\theta$ represents the crack direction and the latter is perpendicular to the vector $\nabla d$ (see Eq.~\ref{theta}). Parameters $\kappa_m$, $\alpha_m$ and $\theta'_m$ are model input parameters. To be able to compare it to the case of weak anisotropy using a second-order structural tensor, we will particularly consider only one energy frequency ($m=1$, $\kappa_1=G_{c,0}$, $\alpha_1=\alpha$ and $\theta'_1=-\phi$). The simplified version of the crack-free energy is written as:
\begin{equation} 
  \label{psiarbit}
\psi_{c,a} =  G_c(\theta)~\gamma_a = G_{c,0} \left( 1+\alpha~\sin^2(\theta-\phi)\right) \left(\dfrac{1}{\pi l_c}~\omega(d) + \dfrac{l_c}{\pi}~||\nabla d||^2 \right).
\end{equation}
Interestingly enough, there are similarities between the current methodology and the modification for the anisotropic crack density function introduced by \cite{YIN2020113202}. Eq.~\ref{psiarbit} shares a lot of similarities with the expression in Eq.~\ref{anisostpsic}, although they are not exactly the same.

Since the elastic part of the energy remains as before, the definition for the stress tensor and damage driving force is the same as described in Eq.~\ref{stressconjugate} and Eq.~\ref{damagedrivingforce}, respectively. Therefore, using the Euler-Lagrange procedure, the variational derivative of the total energy with respect to the displacement field results in the same expression described in Eq.~\ref{deltau}.
Based on the crack density function with arbitrary direction-dependent fracture energy, and considering $G_c(\theta)=g_c(\nabla d)$, for the variational derivative with respect to the damage field we have \cite{REZAEI2021a}
\begin{equation}
\label{aransio} 
\delta_d \psi=0 \Rightarrow \begin{cases}
\dfrac{g_c(\nabla d)}{l_c \pi} \omega^{\prime}-\text{div}\left(\dfrac{l_c~g_c(\nabla d)}{c_0} ~\nabla d\right)-\text{div}(\gamma \bm{g}_d) -\text{div}( \bm{s}_d\,H_a)-Y_{m,a}=0 ~\text{in}~\Omega \\
\nabla d \cdot \bm{n}=0~\text{on}~\Gamma_c
\end{cases}
\end{equation}
Similar to Eq.~\ref{Y_m}, the expression for $Y_{m,a}$ is defined as maximum value between the undamaged elastic strain through the simulation time ($\psi_e^0(t)$) and the new damage energy threshold ($\psi_{th,a}$):
\begin{equation}
  \label{Y_m_a}
-Y_{m,a} = f_D^{\prime}H_a = f_D^{\prime} \max_t (\psi_e^0(t),\psi_{th,a}).
\end{equation}
We choose the following definition for damage threshold (see also \cite{MIEHE2015449}):
\begin{equation}
\label{the_a}
\psi_{th,a}=\dfrac{1}{2E}~\sigma_{u}^2(\theta).
\end{equation}
For the direction-dependent tensile strength $\sigma_{u}(\theta)$, we propose the following function:
\begin{equation}
\label{sigma_u_theta}
\sigma_{u}(\theta) = \sum_m \sigma_{0,m}\left(1+\alpha_{m}~\sin^2\left(m(\theta+\theta'_{m} \right))\right)^{p_m}, ~m\in \mathbf{N}.
\end{equation}
Similar to the direction-dependent fracture energy, here $m$ denotes the frequency number. The total strength of the material is the summation over all the active frequencies. Furthermore, $p_m$ denotes an additional material parameter in this work. The structure of Eq.~\ref{sigma_u_theta} is also motivated by the work of \cite{HOSSAIN2018118} and certainly can be modified according to specific application.
Utilizing Eq.~\ref{sigma_u_theta} allows us to have a directional maximum tensile strength. It worth mentioning that the other parameters such as $\alpha_m$ and $\theta'_m$, are the same as the ones in Eq.~\ref{gcem}.

For the case of arbitrary direction-dependent fracture energy, the following relations are proposed to obtain the constants $a_1$ and $a_2$ in the damage function $f_D$ (see Eqs.~\ref{f_D} and \ref{fdprime}). 
\begin{equation}
  \label{a1_araniso}
a_{1,a}=\dfrac{4 E\,G_c(\theta)}{\pi l_c ~\sigma_{u}^2(\theta)},~~~~~
a_{2,a} = -0.5.
\end{equation}

The two new terms in Eq.~\ref{aransio}, $\bm{g}_d$ and $\bm{s}_d$, are imposed by the directional dependency of the fracture energy function and the degradation function, respectively (compare Eq.~\ref{aransio} to Eq.~\ref{deltadsttensor} and see \cite{REZAEI2021a}).
Finally, we have the following definitions for the new terms in Eq.~\ref{aransio}:
\begin{align}
\label{gd_eq}
\bm{g}_d &=\dfrac{\partial G_c (\theta)}{\partial \nabla d} =  \dfrac{\partial G_c(\theta)}{\partial \theta}~\dfrac{\partial \theta}{\partial \nabla d}, \\[1em]
\label{sd_eq}
\bm{s}_d &=\dfrac{\partial f_D}{\partial \nabla d} =  \dfrac{\partial f_D}{\partial a_1}~\dfrac{\partial a_1}{\partial \theta}~\dfrac{\partial \theta}{\partial \nabla d}.
\end{align}
For the calculation of new terms $\bm{g}_d$ and $\bm{s}_d$ in Eq.~\ref{aransio}, the following steps have to be taken:
\begin{align}
\dfrac{\partial G_c(\theta)}{\partial \theta} &= G_{c,0}\,\alpha\,m\sin(2m (\theta\,+\theta^{\prime})), \\[1em]
\dfrac{\partial f_D}{\partial a_1} &= \dfrac{(1-d)^2 (d-d^2/2)}{\left[(1-d)^2 + a_1 d+a_1 a_2 d^2 \right]^2},\\[1em]
\dfrac{\partial a_1}{\partial \theta} &=\dfrac{4 E\,G_c}{\pi\,l_c\,\sigma_0^2}~\dfrac{\alpha\,m (1-2\,p_m)\,\sin(2m\,(\theta\,+\theta^{\prime}))}{\left(1 + \alpha\,\sin^2(\theta +\theta^{\prime}) \right)^{2p_m}}, \\[1em]
\dfrac{\partial \theta}{\partial \nabla d}&=\dfrac{1}{||\nabla d||^2} \begin{bmatrix} -\nabla d \cdot \bm{e}_2 \\ \nabla d \cdot \bm{e}_1
\end{bmatrix}.
\end{align}

\textbf{Remark 4.} The new terms mentioned in Eq.~\ref{gd_eq} and Eq.~\ref{sd_eq} are the contributions by considering an arbitrary shape for the direction-dependent fracture energy function as strength. As we will discuss in the next section, these terms can be computed explicitly within the finite-element calculation to reduce the complexity of the implementation (see also \cite{REZAEI2021a} and Algorithm 1).

\subsection{Explanation of damage threshold}
Having a linear term in the crack topology function $\omega(d)$ enables us to have an initial elastic stage before damage initiation. In other words, by considering the threshold, one can guarantee that the value of damage remains zero~($d=0$) in Eq.~\ref{deltadsttensor}. 
A simple one-dimensional analysis is carried out for clarification. Considering a uniform distribution for the damage variable ($d'=\partial d/\partial x=0$), the governing equation for damage (Eq.~\ref{deltadsttensor}) reduces to
\begin{equation}
\label{thresholdeq}
\dfrac{G_c}{l_c \pi} (2-2d)-f_D^{\prime} H=0.
\end{equation}
Note that if there is no damage threshold $\psi_{th}$, damage takes the value one. Considering Eqs.~\ref{Y_m} and \ref{the_a}, one can further simplify the damage governing equation to
\begin{equation}
  \label{thresholdeq1}
\dfrac{G_c}{l_c \pi} (2-2d)-a_1 \dfrac{1}{2E} \sigma_0^2=0.
\end{equation}
Having Eq.~\ref{a1_st_tensor} in hand, the above expression guarantees that the damage value remains $0$ before the threshold is met. 
After passing the threshold (i.e. $\psi_e^0>\psi_{th}$), the history parameter $H$ in Eq.~\ref{thresholdeq} is replaced by $\psi^0_e = \dfrac{1}{2} \bm{\epsilon}:\mathbb{C}_h:\bm{\epsilon}$ which derives the damage to evolve.

\subsection{Cohesive zone model}
Here we summarize the formulation of the cohesive zone model (CZM). The CZM relates the traction vector $\boldsymbol{t}=[t_n,~t_s]^T$ to the displacement jump or gap vector $\boldsymbol{g}=[g_n,~g_s]^T$:
\begin{align}
{t}_n &= k_0\left(1-D\right){g}_n,\\
\label{eq:cz_biliear}
{t}_s &= \beta^2 k_0\left(1-D\right){g}_s.
\end{align}
Here, $k_0$ is the initial stiffness of the cohesive zone model. Damage at the interface ($D$) is determined based on the introduced traction-separation relation \cite{Rezaei2017, Khaledi2018b}:
\begin{equation}
\label{eq:czdamage}
   D = \displaystyle \begin{cases} \displaystyle 0 & \mbox{if } \lambda < \lambda_0 \\
\displaystyle \frac{\lambda_f}{\lambda_f-\lambda_0}\frac{\lambda-\lambda_0}{\lambda} & \mbox{if } \lambda_0 < \lambda < \lambda_f \\ \displaystyle 1 & \mbox{if } \lambda_f<\lambda \end{cases}.
\end{equation}
The parameter $\lambda = \sqrt{\langle{g}_n\rangle^2 + (\beta{g}_s)^2}$ represents the amount of separation at the interface with ${g}_n$ and ${g}_s$ being the normal and shear gap vector, respectively. The parameters of the model are summarized as (1) the maximum strength of the interface $t_0= k_0\lambda_0$, (2) the critical separation where damage starts $\lambda_0$, (3) the final separation at which the traction goes to zero $\lambda_f$, and (4) the parameter $\beta$ which governs the contribution of the separation in shear direction. As a result, the interface fracture energy is computed using $G_{c,int} = \dfrac{1}{2} t_0 \lambda_f$.


\subsection{Summary of different formulations}
Here we would like to compare different formulations for the convenience of the reader. First off, we have the comparison between modeling anisotropic fracture utilizing structural tensor and arbitrary direction-dependent fracture energy in Table~\ref{tab:svsa}. Note that both of these anisotropic formulations are based on cohesive fracture models \cite{LORENTZ201120,wu2018,GEELEN2019}.

For the sake of completeness, a comparison is also performed between the standard PF damage model and cohesive PF damage model in Table~\ref{tab:svsc}. The standard phase-field model which is used in this work is based on the so-called AT-2 (see also \cite{Miehe2010a}). Readers are also encouraged to see Appendix A and B. The formulations in Table~\ref{tab:svsc} can be simply coupled with those in Table~\ref{tab:svsa} to construct anisotropic cohesive phase-field models.
\begin{table}[H]
    \centering
    \begin{tabular}{@{} llll @{}} 
    \toprule
                 & 
    Structural tensors      & 
    Arbitrary fracture energy function    \\ 
    \midrule
    Crack energy & 
    $\psi_{c,s} = G_{c,0}~\gamma_s(d,\nabla d)$
    & 
    $\psi_{c,a} = G_c(\theta)~\gamma_a(d,\nabla d)$& 
    \\
                 & 
    $\gamma_s = \dfrac{\omega(d)}{c_0 l_c}+\dfrac{l_c}{c_0}~\nabla d \cdot \bm{A} \cdot \nabla d$ & 
    $\gamma_a = \dfrac{\omega(d)}{c_0 l_c}+\dfrac{l_c}{c_0}~\nabla d \cdot \nabla d$ & 
    \\
    &
    $G_{c,0} = \text{const.}$ &
    $G_c(\theta) = \sum_m \kappa_m\left(1+\alpha_m \sin^2\left(m(\theta+\theta'_m \right))\right)$ 
    \\ \\
    Damage function & 
    $f_D = \dfrac{(1-d)^2}{(1-d)^2+a_1d+a_1a_2d^2}$ & 
    $f_D = \dfrac{(1-d)^2}{(1-d)^2+a_1d+a_1a_2d^2}$ & 
    \\
                  &  
    $a_{1,s} = (4 E G_c)/(\pi l_c ~\sigma_{0}^2)$ & 
    $a_{1,a}=(4 E G_{c}(\theta))/(\pi l_c ~\sigma_{u}^2(\theta))$  & 
    \\
      &
    $a_{2,s} = -0.5$ 
    &
    $a_{2,s} = -0.5$
    \\
                 & 
    $\sigma_{0} = \text{const.}$            & 
    $\sigma_{u}(\theta) = \sum_m \sigma_{0,m}\left(1+\alpha_{m} \sin^2\left(m(\theta+\theta'_{m} \right))\right)^{p_m}$   &
    \\ \\
    Damage threshold        & 
    $\psi_{th,s} = \dfrac{1}{2E}\sigma_0^2$ 
                  & 
    $\psi_{th,a} = \dfrac{1}{2E}~\sigma_{u}^2(\theta)$
    \\
    \bottomrule
    \end{tabular}
	\caption{Comparison between anisotropic damage models based on structural tensors and arbitrary fracture energy function.}
	\label{tab:svsa}
\end{table}

\begin{table}[H]
    \centering
    \begin{tabular}{@{} llll @{}} 
    \toprule
                 & 
    Standard phase-field model         & 
    Cohesive phase-field model    \\ 
    \midrule
    Crack topology function & 
    $\omega(d) = d^2$ & 
    $\omega(d) = 2d - d^2$ 
    \\
                 & 
    $c_0=2$ & 
    $c_0=\pi$ 
    \\ \\
    Damage function & 
    $f_D = (1-d)^2$ & 
    $f_D = \dfrac{(1-d)^2}{(1-d)^2+a_1d+a_1a_2d^2}$ 
    \\
                  &  
    --- & 
    $a_{1,s} = (4 E G_c)/(\pi l_c ~\sigma_{0}^2),~~
a_{2,s} = -0.5$  
    \\ \\
    History parameter & 
    $H=\max_t (\psi_e^0(t))$ & 
    $H=\max_t (\psi_e^0(t),\psi_{th})$ & 
    \\
                  &  
    --- & 
    $\psi_{th} = \dfrac{1}{2E}\sigma_0^2$  
    \\
    \bottomrule
    \end{tabular}
	\caption{Standard versus cohesive phase-field damage models.}
	\label{tab:svsc}
\end{table}

\section{Weak form and discretization} 
Through the FE discretization procedure, the following approximation for displacement and damage fields within a typical element and their derivatives are employed (see \cite{Brepols2017,RDeborst2016}). 
\begin{equation}
\label{discrz}
\begin{cases}
\bm{u}=\bm{N}_u \bm{u}_e \\
d = \bm{N}_d \bm{d}_e
\end{cases},
~~~~~~~
\begin{cases}
\bm{\epsilon}=\bm{B}_u \bm{u}_e \\
\nabla d = \bm{B}_d \bm{d}_e 
\end{cases}.
\end{equation}
The subscript $e$ represents the nodal values of the corresponding quantity. Utilizing linear shape functions and considering a quadrilateral $2$D element, one obtains the following matrices for shape functions and their derivatives in $\bm{N}$~and~$\bm{B}$ matrices, respectively:
\begin{equation}
\label{Nmatrix}
\bm{N}_u=
\begin{bmatrix}
N_1 & 0 & \cdots & N_4 & 0\\
0 & N_1 & \cdots & 0 & N_4
\end{bmatrix}_{2\times8},
~~~~~~\bm{N}_d=
\begin{bmatrix}
N_1 & N_2 & N_3 & N_4 
\end{bmatrix}_{1\times4},
\end{equation}

\begin{equation}
\label{Bmatrix}
\bm{B}_u=
\begin{bmatrix}
N_{1,x} & 0 & \cdots & N_{4,x} & 0\\
0 & N_{1,y} & \cdots & 0 & N_{4,y} \\
 N_{1,y} &N_{1,x}& \cdots & N_{4,x} & N_{4,y}
\end{bmatrix}_{3\times8},
~~~\bm{B}_d=
\begin{bmatrix}
N_{1,x} & N_{2,x} & N_{3,x} & N_{4,x}\\
N_{1,y} & N_{2,y} & N_{3,y} & N_{4,y}
\end{bmatrix}_{2\times4}.
\end{equation}

Next, the weak form of the governing Eqs~\ref{deltau}, \ref{deltadsttensor} and \ref{aransio} are obtained. After applying the introduced discretization form, the residual vectors for the Newton-Raphson solver are obtained for the displacement and damage field:
\begin{equation}
\label{ru}
\bm{r}_u=-\left[ \left( \int_{\Omega_e} \bm{B}^T_u \bm{C} \bm{B}_u \bm{u}_e - \bm{N}^T_u \bm{b} \right) dV - \int_{\partial \Gamma_t} \bm{N}^T_u \bm{t}~dA \right]_{8\times1}
\end{equation}

\begin{equation}  
\label{rds}
\bm{r}_{d,s}=-\left[ \int_{\Omega_e} \dfrac{2G_{c,0} l_c}{\pi} \bm{B}^T_d \bm{A}\bm{B}_d \bm{d}_e +\bm{N}^T_d\left(\dfrac{\omega^{\prime}(d)G_{c,0}}{c_0l_c} - Y_{m,s} \right) dV \right]_{4\times1} 
\end{equation}

\begin{equation}
\label{rda}  
\bm{r}_{d,a}=-\left[ \int_{\Omega_e} \dfrac{2 g_c(\nabla d) l_c}{\pi} \bm{B}^T_d \bm{B}_d \bm{d}_e +\bm{N}^T_d\left(\dfrac{\omega^{\prime}(d) g_c(\nabla d)}{c_0l_c} - Y_{m,a} \,+\,\gamma \,\bm{B}^T_d\,\bm{g}_d\,+H_a^i\,\bm{B}^T_d\,\bm{s}_d\right) dV \right]_{4\times1} 
\end{equation}
The above equation are the element residuals. The residual vector for the case of structural and arbitrary anisotropy are denoted by $\bm{r}_{d,s}$ and $\bm{r}_{d,a}$, respectively. Note that either $\bm{r}_{d,s}$ or $\bm{r}_{d,a}$ is used in what follows. As shown in Algorithm~1, we utilize a staggered approach which is known to be able to handle the instabilities upon damage progression in a more robust way \cite{GERASIMOV2016276,Zhang201818}.
The superscript emphasizes the semi-implicit scheme which is used for computing the direction of the crack. The residuals and stiffness matrix at the element level are shown in Algorithm~1. Here, the solver finds the solution at time $i+1$ using an iterative approach till $\Delta\boldsymbol{u}_{e,k+1}^{i+1}=\Delta\boldsymbol{ D}_{e,k+1}^{i+1}=\boldsymbol{0}$. The letter $k$ represents the Newton iteration number.
\begin{align}
\begin{bmatrix}
\Delta\boldsymbol{u}_{k+1}^{i+1} \\
\Delta\boldsymbol{d}_{k+1}^{i+1} 
\end{bmatrix} =
\begin{bmatrix}
\boldsymbol{u}_{k+1}^{i+1}-\boldsymbol{u}_{k}^{i+1} \\
\boldsymbol{d}_{k+1}^{i+1}-\boldsymbol{d}_{k}^{i+1} 
\end{bmatrix} =
-
\begin{bmatrix}
\boldsymbol{K}^{i+1}_{uu,k+1} & 0  \\
0 & \boldsymbol{K}^{i+1}_{dd,k+1} 
\end{bmatrix}^{-1}
\begin{bmatrix}
\boldsymbol{R}^{i+1}_{u,k+1} \\
\boldsymbol{R}^{i+1}_{d,k+1} 
\end{bmatrix}.
\end{align}
In the above equation, $\boldsymbol{R}^{i+1}$ and $\boldsymbol{K}^{i+1}$ denote the assembled global residual vector and stiffness matrices, respectively.
\begin{algorithm}[H]
\label{alg:ArMainAlgo}
\SetAlgoLined 
\caption{Element residual vector and stiffness matrix for the case of arbitrary anisotropy}
Inputs: ~~$\bm{d}_e^i$, $\bm{u}_e^i$, $||\nabla d_c||$ and material properties \\ 
Outputs: $\bm{d}_e^{i+1}$~and  $\bm{u}_e^{i+1}$ \\
$\nabla d^i =\bm{B}_d \bm{d}_e^i$ $\to$    $\theta^{i} =\text{tan}^{-1} \left( \dfrac{\nabla d^i_y}{\nabla d^i_x} \right)$\\
  \eIf{$||\nabla d^i|| \ge ||\nabla d||_c$}{
   Compute $g_c^{i}$ (Eq.~\ref{gcem}) and $\sigma_{u}^{i}$ (Eq.~\ref{sigma_u_theta}) \\
   Compute $\bm{g}^i_d$ (Eq.~\ref{gd_eq}) and $\bm{s}^i_d$ and (Eq.~\ref{sd_eq})
   }{
   Set $g_c^i = G_{c,min}$ and $\sigma_{u}^{i} ={\sigma_{u,min}}$  \\
   Set $\bm{g}^i_d=\bm{0}$ and $\bm{s}^i_d=\bm{0}$ 
  } 
 Compute $\psi_e^{i+1}=\dfrac{1}{2}\bm{\epsilon}^{i+1}:\mathbb{C}^i:\bm{\epsilon}^{i+1}$ (Eq.~\ref{elasticenergy}) and $\psi_{th,a}^{i}$ (Eq.~\ref{the_a})
 
 Compute $H_a^{i}= \max (\psi_e^{i},\psi_e^{i+1}, \psi^{i}_{th,a})$ (Eq.~\ref{Y_m_a}) and $a^i_{1,a}$ (Eq.~\ref{a1_araniso})
 \\ \\
 ${\bm{r}_{u}^{i+1}}= \displaystyle\int_{\Omega_e} \left(\bm{B}^T_u \bm{C}^{i} \bm{B}_u {\bm{u}_{e}^{i+1}} - \bm{N}^T_u \bm{b}  \right) dV + \displaystyle\int_{ \Gamma_t} \bm{N}^T_u \bm{t}~dA$
 \\ 
  $\bm{r}_{d}^{i+1}=\displaystyle\int_{\Omega_e} \left(\dfrac{2 g_c^{i} l_c}{\pi} \bm{B}^T_d \bm{B}_d {\bm{d}_{e}^{i+1}} +\bm{N}^T_d\left(\dfrac{\omega^{\prime}({\bm{d}_{e}^{i+1}}) g_c^{i}}{\pi l_c} + f_D^{\prime}(\bm{d}_{e}^{i+1})H_a^i \right) \,+\,\gamma \,\bm{B}^T_d\,\bm{g}^i_d\,+H_a^i\,\bm{B}^T_d\,\bm{s}^i_d\right) dV$  
\\ 
  ${\bm{k}_{uu}^{i+1}}= \displaystyle\int_{\Omega_e}  \bm{B}^T_u \bm{C}^{i} \bm{B}_u ~dV$
\\ 
  ${\bm{k}_{dd}^{i+1}} =\displaystyle\int_{\Omega_e} \left( \dfrac{2l_c g_c^{i}}{\pi}\bm{B}^T_d \bm{B}_d +  \bm{N}^T_d \left( \dfrac{\omega^{\prime \prime}({\bm{d}_{e}^{i+1}}) g_c^{i} }{\pi l_c} + f_D^{\prime \prime}(\bm{d}_{e}^{i+1})H_a^{i}  \right) \bm{N} \right)dV$\\ 
\end{algorithm}
Algorithm 1 is written at the element level. As discussed in \cite{REZAEI2021a}, the parameter $||\nabla d||_c$ is introduced since at the beginning of the simulation there is no damage to determine the direction-dependent property based on $\nabla d$. According to studies in \cite{REZAEI2021a}, its value should be large enough for avoiding convergence issues. We will provide suggestion for choosing this parameter it what follows. Note that the explicit evaluation of $\bm{g}_d$~and~$\bm{s}_d$, causes the vanishing of these terms in element stiffness.
\textcolor{green}{}
%

\section{Numerical examples}
The material parameters used for the following numerical studies are reported in Table~\ref{tab:par_mat}. We will focus on damage propagation in an elastic solid with an initial crack. Note that for the first set of studies, the elastic constants are not rotated according to the preferential fracture direction (i.e. we have initially isotropic material). The anisotropic elastic properties will influence the crack direction as well (see \cite{ZHANG2020112643} for such studies). By focusing on initially isotropic material, one can better focus on the influence of the direction-dependent fracture energy on the crack path. Further studies on the combined influence of anisotropic elasticity and fracture are postponed to future studies.  
\begin{table}[H]
	\centering
	\begin{tabular}{ l l l } \hline
	\multirow{1}{*}{}         & Unit    & Value    \\ \hline \hline 
	Lamé's Constants ($\lambda,~\mu$)   & [GPa] & $(132.6, 163.4)$  \\
	Fracture energy $G_c$ $=G_{c,0}$  & [$\dfrac{\text{J}}{\text{m}^2}$] $\equiv$ [GPa.$\mu \text{m}$]$10^3$ & $40$\\
	Ultimate strength $\sigma_0$ $=\sigma_{0,1}$  & [GPa] & $5$\\
	Damage internal length $l_c$    & [$\mu$m]  & $0.025-0.2$   \\ 
	Frequency number $m$            & [-]   & $1,~2$  \\ 
	Fracture energy parameter $\alpha_m$     & [-] & $0.0,~3.0$ \\ 
	Fracture energy parameter $\theta'_m$    & [-] & $-40^\circ,~0^\circ$       \\
    Structural parameter $\alpha$    & [-] & $0.0 , 12.0$ \\
    Structural parameter $\phi$    & [-] & $-40^\circ,~0^\circ$  \\
    Damage parameter $||\nabla d||_c$  & [-] & $0.2$ \\
    Material strength parameter $p_m$ & [-] & $0.1$    \\
    \hline
	\end{tabular}
	\caption{Parameters for the anisotropic PF damage formulation.}
	\label{tab:par_mat}
\end{table}


\subsection{Crack propagation in an initially isotropic solid}
\begin{figure}[H]
    \centering
    \includegraphics[width=1.0\linewidth]{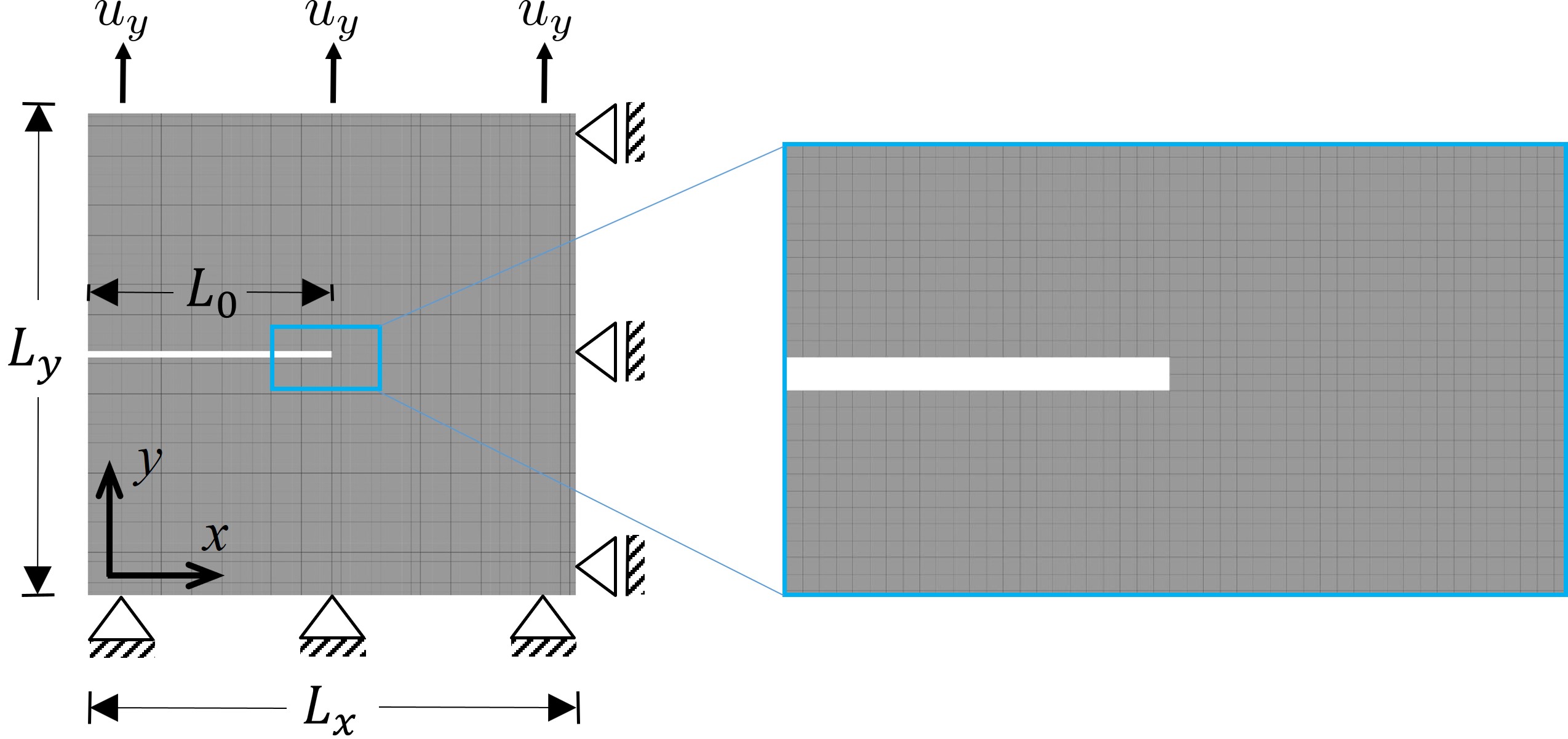}
    \caption{Boundary conditions and geometry of a single notched specimen.}
    \label{fig:geom}
\end{figure}

According to Fig.~\ref{fig:geom}, a single notched specimen is studied. Simulations are carried out in a 2D configuration based on a plane-strain assumption. Two different dimensions are chosen for numerical studies (see Table~\ref{tab:par_geo}). Geometry B is constructed by scaling geometry A by the factor $1/8$. As will be shown, choosing the smaller geometry will help us to motivate and understand better the idea behind cohesive fracture. Moreover, on the right-hand side of Fig.~\ref{fig:geom}, the mesh topology is illustrated. In all simulations, we make sure that enough elements are utilized depending on the chosen value for the length scale parameter $l_c$. 

\begin{table}[H]
	\centering
	\begin{tabular}{ l l l } \hline
	\multirow{1}{*}{}         & geometry A    & geometry B    \\ \hline \hline 
	Length in $x$ direction $L_{x}$, [$\mu$m] &$4.0$ & $0.5$  \\
	Length in $y$ direction $L_{y}$, [$\mu$m] &$4.0$ & $0.5$  \\
	Initial crack length $L_{0}$,    [$\mu$m] &$2.0$ & $0.25$ \\
	 \hline
	\end{tabular}
	\caption{Chosen dimensions for the numerical studies. The geometry A is $8$ times larger than the geometry B.}
	\label{tab:par_geo}
\end{table}

We will start by assuming a constant fracture energy value also known as isotropic crack propagation. For the defined boundary value problem in Fig.~\ref{fig:geom}, the crack propagates along a horizontal line without any deviation. The system with geometry A is simulated utilizing different models. 

\begin{figure}[H]
    \centering
    \includegraphics[width=0.9\linewidth]{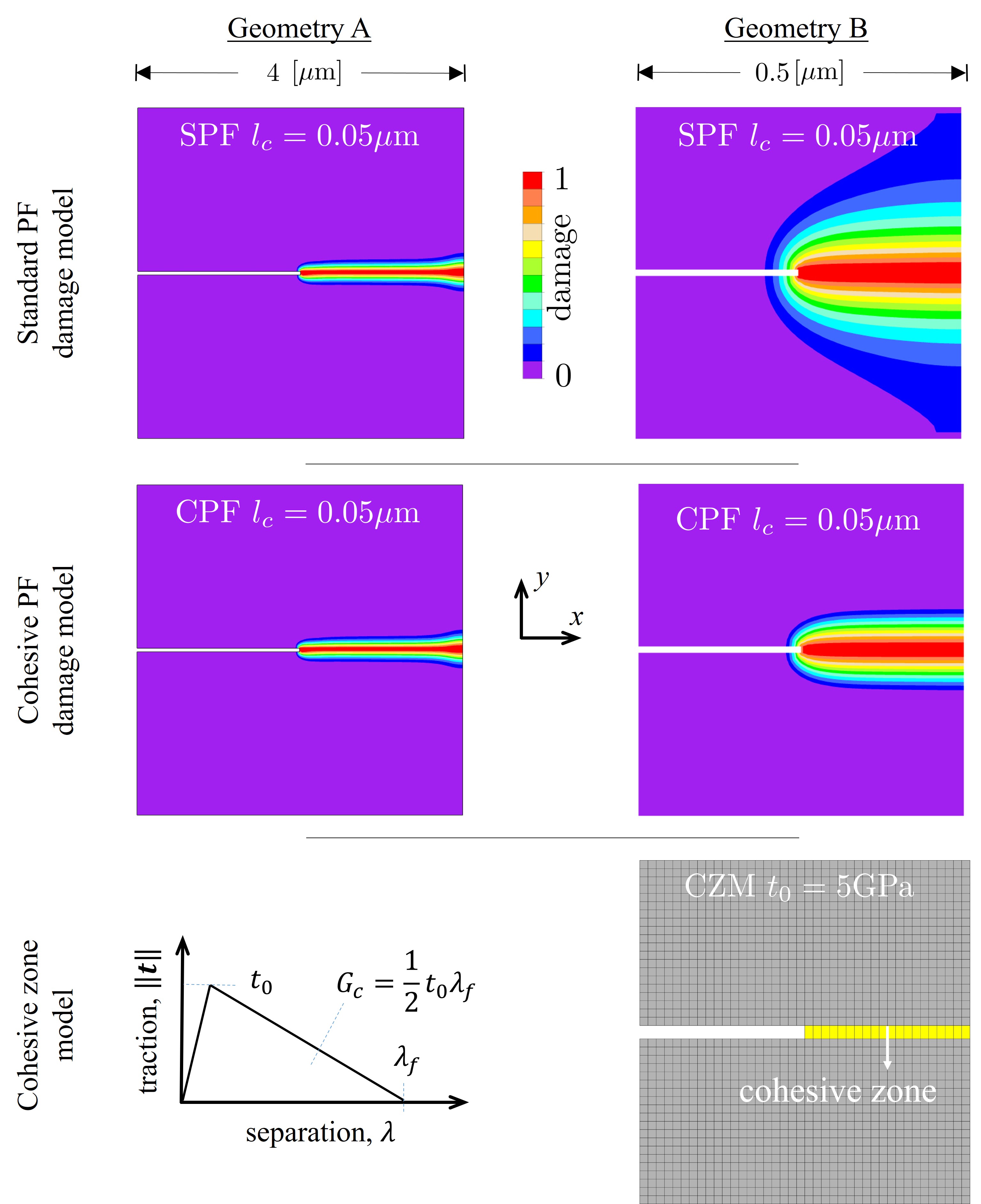}
    \caption{comparing the crack paths of CPF, SPF and CZM models (for the isotropic damage case)    }
    \label{fig:LvsS}
\end{figure}

In Fig.~\ref{fig:LvsS}, the results of the standard phase-field (SPF) damage, cohesive phase-field (CPF) damage as well as cohesive zone (CZ) model are presented in different rows (see also Table~\ref{tab:abb}). 
In the simulation using the SPF damage model, the internal length scale parameter is set to $l_c=0.05~\mu$m. The latter value is chosen based on the available analytical relations between the internal length scale and other material properties (see \cite{TenneBourdin}):,
\begin{equation}
\label{lclc}
\sigma_0 = \dfrac{9}{16}\sqrt{\dfrac{E\,G_c}{3\,l_c}} \Rightarrow l_c \approx 0.05~\mu m.
\end{equation}

\begin{table}[H]
	\centering
	\begin{tabular}{ l l l } \hline
	\multirow{1}{*}{}  Abbreviation    & Model    \\ \hline \hline 
	SPF & Standard Phase-Field \\
	CPF & Cohesive Phase-Field \\
	CZM & Cohesive Zone Model  \\
	 \hline
	\end{tabular}
	\caption{Summary of the models utilized in this work.}
	\label{tab:abb}
\end{table}

In the last row of Fig.~\ref{fig:LvsS}, the same boundary value problem is calculated utilizing the standard bi-linear CZ model \cite{Rezaei2017}. Since we know that the crack propagates in the horizontal direction, CZ elements are introduced accordingly. 

For the first study, the interface behavior is assumed to be isotropic, i.e. $\beta=1$. The CZ parameters such as the maximum strength of the CZ model ($t_0$), the undamaged stiffness of cohesive zone model ($k_0$), and the area beneath the TS curve ($G_{c,int}$) are chosen to represent very similar material properties reported in Table~\ref{tab:par_geo}. Therefore, $\lambda_f=0.016~\mu \text{m}$ is obtained. Moreover, the CZ initial stiffness is set to $k_0=5\times 10^{12}~[\dfrac{\text{GPa}}{\mu \text{m}}]$ to get the closet possible result to the phase-field approach.

Comparing the results obtained from SPF and CPF for the larger geometry does not show any obvious difference. In other words, when the dimension of the problem ($L_x$) is comparatively larger than the internal length scale ($l_c$), the SFP performs well enough. The latter point is well accepted in many engineering applications and, therefore, motivated many researchers to treat the parameter $l_c$ as a material parameter. On the other hand, when it comes to geometry B, utilizing SPF results in a wide spread of the damage zone. Although the same internal length scale parameter is used for the simulation with CPF, the damage zone is much more localized in a certain region (see the idea of the threshold for damage introduced in Section 2.4). 

\textbf{Remark 5.} The spread of the damage zone for the case of the SPF formulation is only problematic, if the geometry is relatively small. One remedy is decreasing the internal length scale which leads to a narrower zone. However, by doing so, we will change the basic material properties that we have (i.e. maximum tensile strength) which is not allowed. Utilizing the CPF formulation, one can select smaller values for $l_c$ depend on the dimension of the problem.

\begin{figure}[H]
    \centering
    \includegraphics[width=1.0\linewidth]{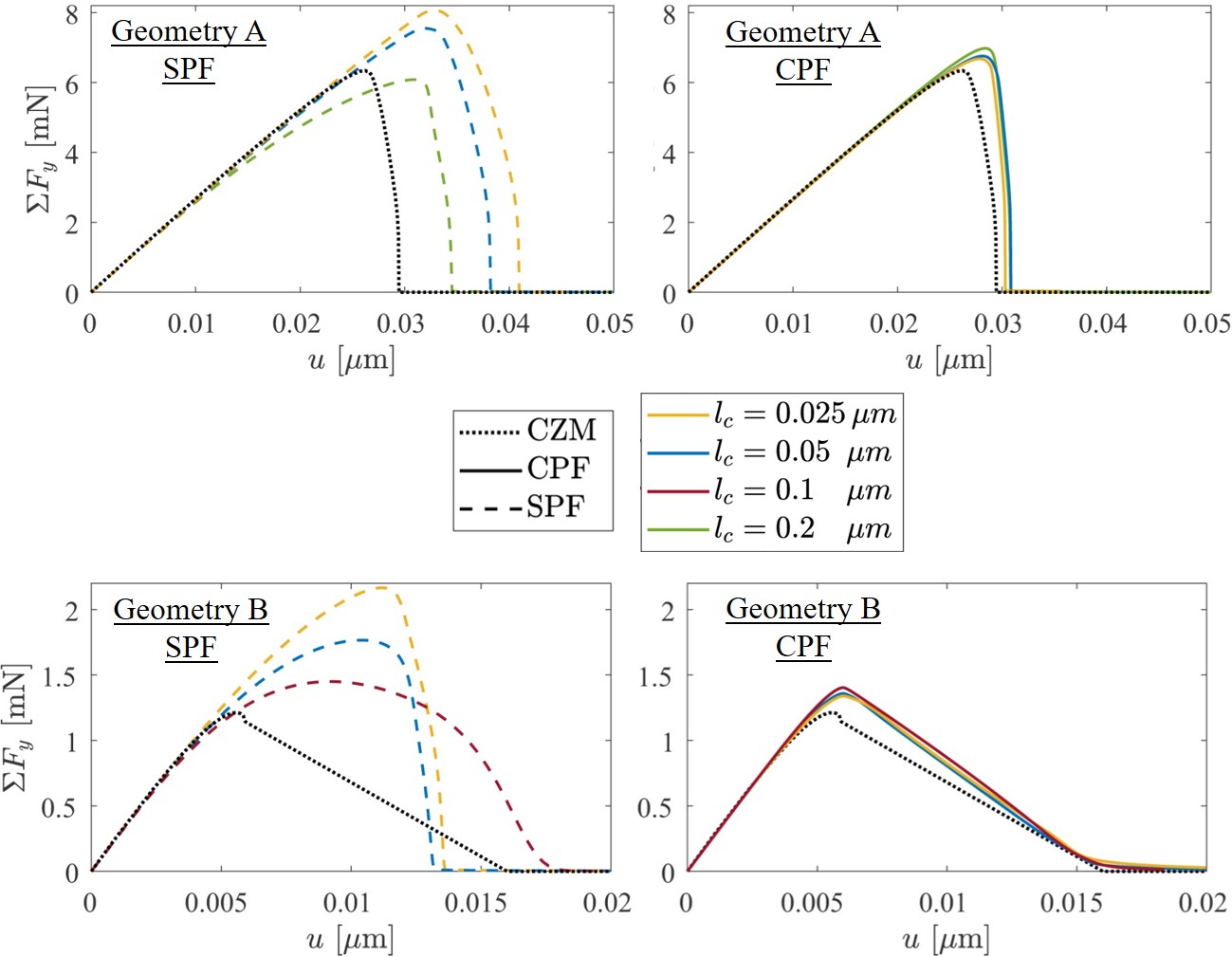}
    \caption{Comparing the response of CPF, SPF and CZM models for the isotropic damage case. The upper row is related to geometry A with larger sizes and the lower one is related to geometry B.}
    \label{fig:RFLvsS}
\end{figure}

The total reaction force obtained from the calculations versus the applied displacement at the top edge is plotted in Fig.~\ref{fig:RFLvsS}. For geometry A, with larger sizes (top row), one observes the typical sharp drop upon sudden and brittle fracture. The results obtained by using the CZ model matches also very well with the SPF models. 
Using SPF and decreasing the parameter $l_c$, the peak point of the reaction force increases as expected. In other words, one can fit the $l_c$ parameter such that the peak point matches well with the results of the CZ model. 
Using the CPF model, the values for the reaction force are almost insensitive with respect to the internal length scale $l_c$ (see also \cite{wu2018} for similar results). This is due to the fact that more information about the fracture property is now included in the model (namely the strength of the material $t_0$ which is not the case for SPF models). Interestingly enough, the results of the CPF model are pretty much following the CZ model which confirms our latter statement.

Focusing on the results of geometry B, one notices a smooth transition in the reaction force after the maximum load peak is reached. The latter observation only holds for the CPF and the CZ models. Note that, we store less elastic energy in geometry B with a smaller size compare to geometry A. In other words, it will be easier for the system to dissipate this total energy by means of crack propagation. Similar to the previous case, the results of the SPF formulation show a clear sensitivity with respect to the length scale parameter $l_c$, while the CPF formulation is not only almost insensitive but also matches very well with the CZ model results. 

\textbf{Remark 6.} Due to size difference, the stored elastic energy is much higher in geometry A than in geometry B. Since the crack surface cannot dissipate all this energy, we observe a sudden drop in the reaction force plots. The sudden drop is due to the staggered algorithm which is used in this study. One may use other techniques like the arc-length method to capture snap-back for geometry A \cite{SINGH2016, Brepols2017, Gibson_Shahed}. Similar behavior is expected using artificial viscous parameter in solving the system of equations \cite{Miehe2010a, Rezaei2017}.

\subsection{Anisotropic crack propagation utilizing structural tensor}
We look at anisotropic cracking in specimens described in Table~\ref{tab:par_geo}, now utilizing the formulation based on the structural tensor $\bm{A}$ (see Eqs.~\ref{anisostpsic}, \ref{deltadsttensor} and \ref{rds}). The model parameters are reported in Table~\ref{tab:par_mat}. Note that by utilizing a structural tensor one can obtain the equivalent fracture energy distribution as a function of the crack angle. Here, by setting $\alpha=12$ the ratio between the maximum and the minimum energy value is equal to $3.0$. This ratio will be used directly in further studies. 
\begin{figure}[H]
    \centering
    \includegraphics[width=1.0\linewidth]{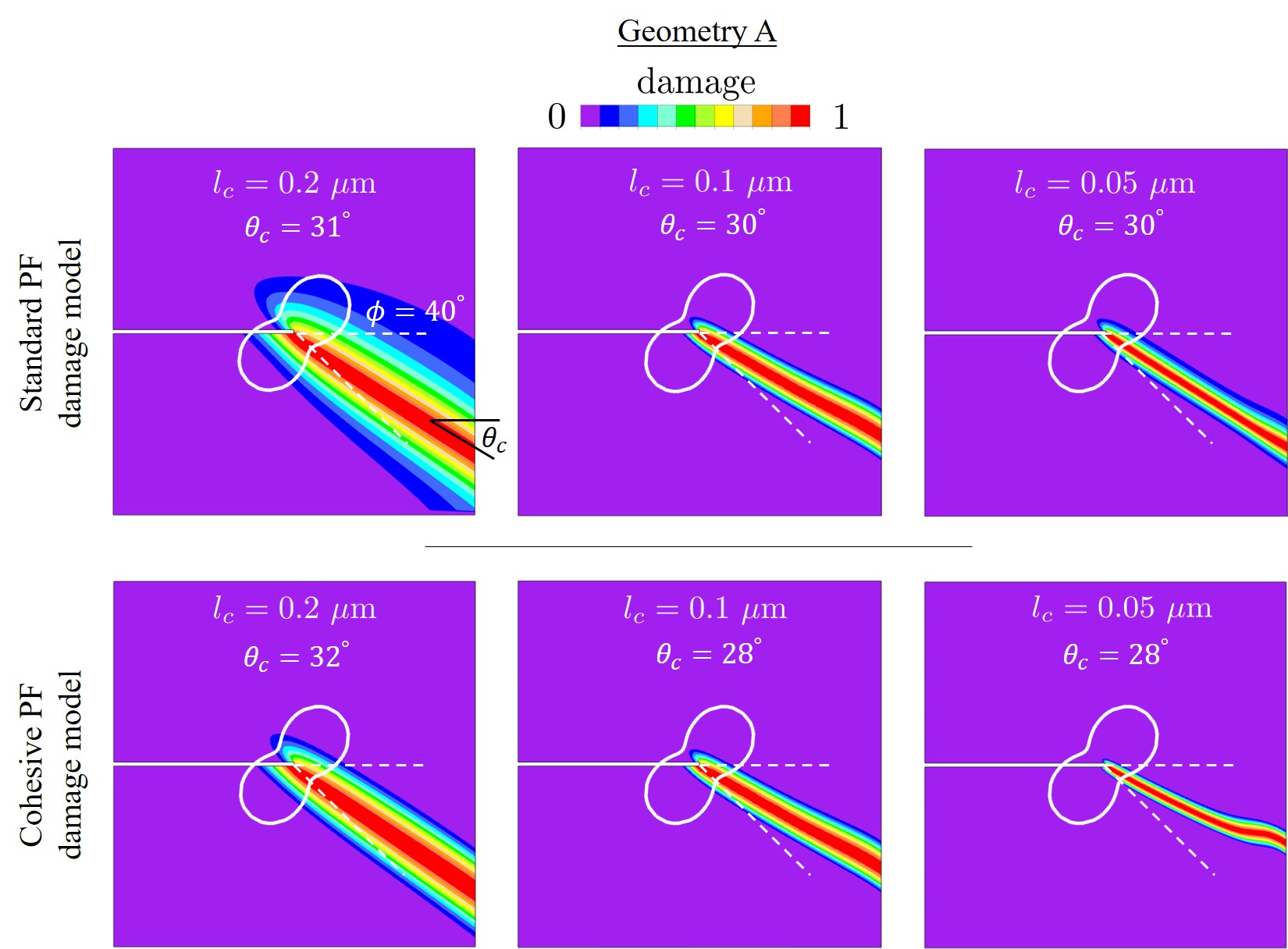}
    \caption{Studies on anisotropic crack propagation using standard and cohesive PF damage formulation utilizing different length scale parameters $l_c$. Here, the geometry A is used.}
    \label{fig:anSCPF_L}
\end{figure}

\begin{figure}[H]
    \centering
    \includegraphics[width=1.0\linewidth]{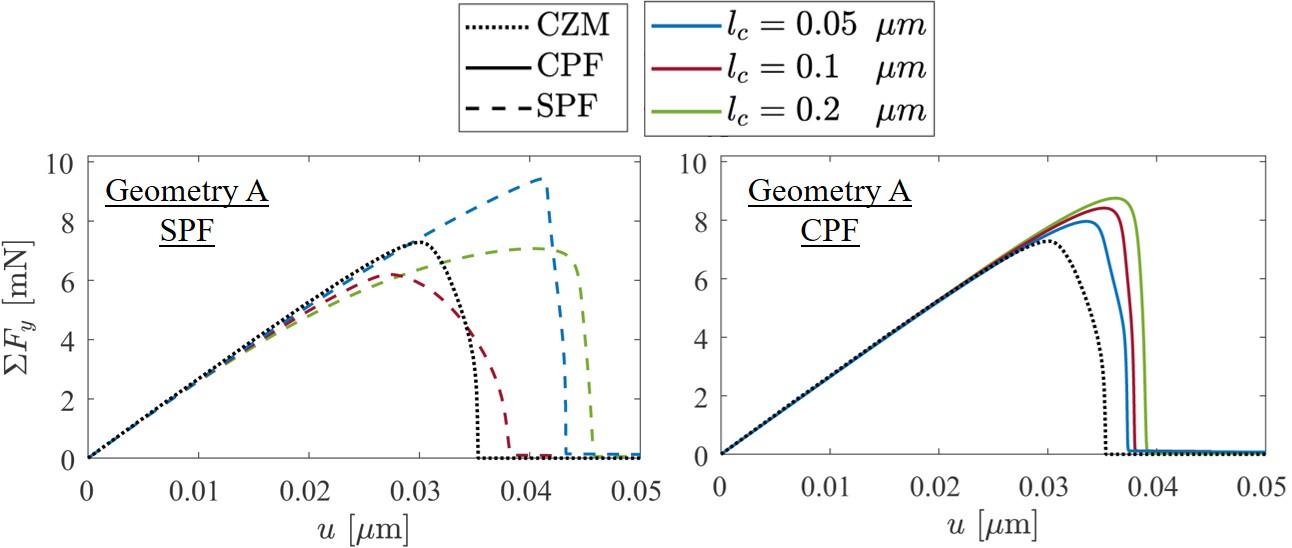}
    \caption{Comparing the response of CPF, SPF and CZM models for the anisotropic damage case.}
    \label{fig:anSCPF_L_RF}
\end{figure}
The fracture energy distribution in the polar coordinate is also plotted in the corresponding figures (see white peanut-shaped curves in Fig.~\ref{fig:anSCPF_L}). According to Fig.~\ref{fig:anSCPF_L}, the influence of the length scale parameter $l_c$ on the obtained crack path is studied. It seems that for both approaches, the crack path angle converges to a certain value $\theta_c \approx 30^{\circ}$. By increasing the parameter $\alpha$, the angle $\theta_c$ converges to the preferential crack direction $\phi$ \cite{REZAEI2021a, MANDAL2020105941, Teichtmeister17}. Similar to the isotropic case, the crack path obtained by SPF and CPF are very close together. Moreover, for a given length scale $l_c$, the damage zone using SPF is relatively wider compared to CPF.

The reaction forces for the aforementioned simulations are shown in Fig.~\ref{fig:anSCPF_L_RF}. For the case of SPF, the reaction forces indicate the dependency of the strength to the length scale parameter. On the other hand, utilizing the CPF, the obtained reaction forces are very much similar with just a slight increase in the peak force. For anisotropic media, one can still use the simplified analytical solutions to relate the maximum strength of the material to the internal length $l_c$. We will try to address this point in the next part.

\begin{figure}[H]
    \centering
    \includegraphics[width=0.9\linewidth]{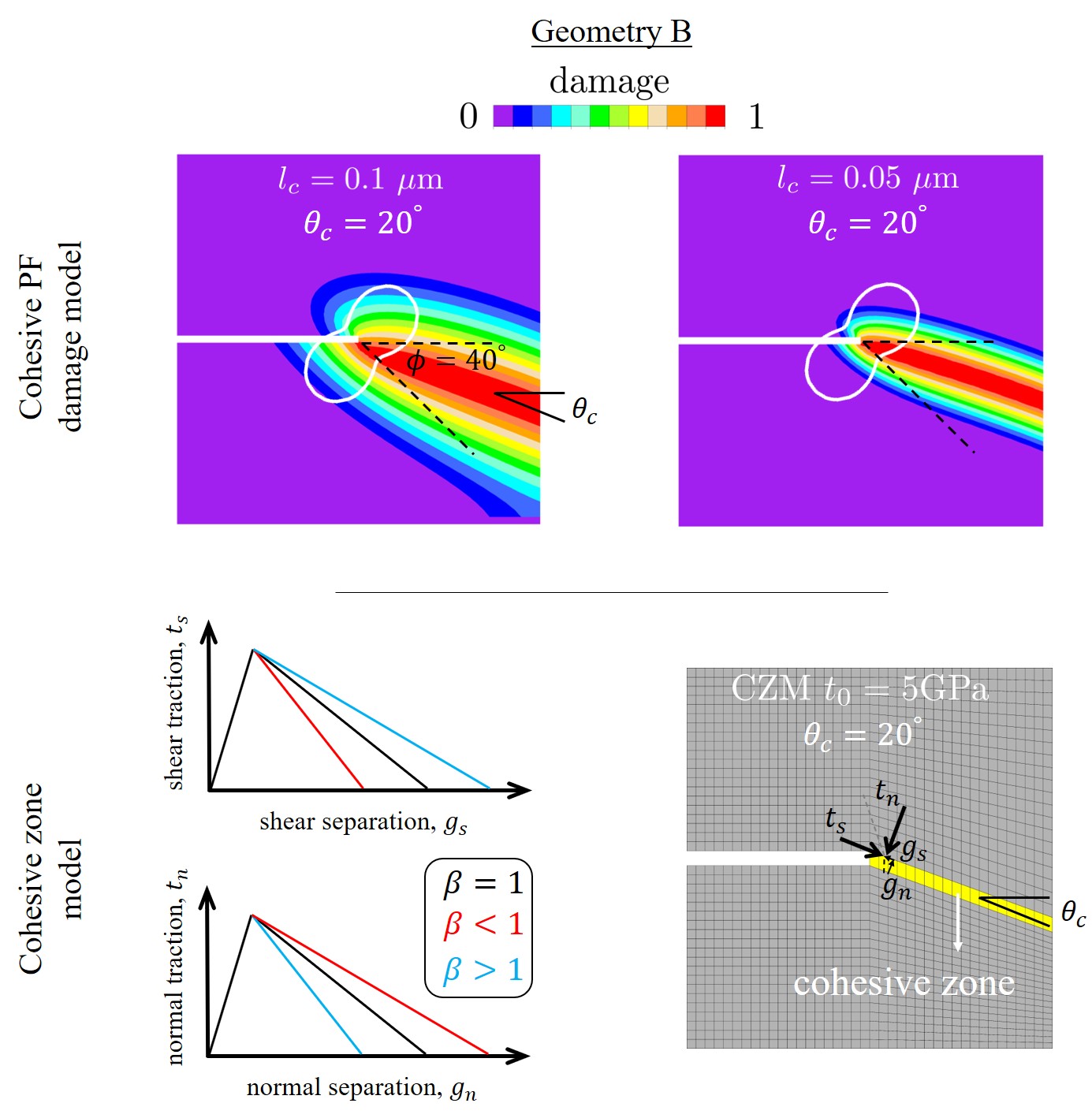}
    \caption{Top: studies on anisotropic crack propagation by using cohesive phase-field damage formulation utilizing different length scale parameter $l_c$. Here, geometry B is used and the material properties are the same for all these studies. Bottom: Simulation results using CZ with different values for $\beta$. }
    \label{fig:anSCPF_S}
\end{figure}

Next, we focus on geometry B, where the specimen dimensions are relatively small and closer to the chosen length scale parameter $l_c$. The crack paths using SPF and CPF are pretty much similar, even for the case of anisotropic fracture energy. Therefore, in Fig.~\ref{fig:anSCPF_S}, only the results of the CPF model are shown. Due to the new geometry dimensions, the final crack path slightly changes to $\theta_c \approx 20^{\circ}$. Note that the material properties such as preferential crack direction $\phi$ are the same as before. Nevertheless, the amount of stored elastic energy and its competition with the crack energy determines the final crack path which is different compared to geometry A.

\begin{figure}[H]
    \centering
    \includegraphics[width=0.8\linewidth]{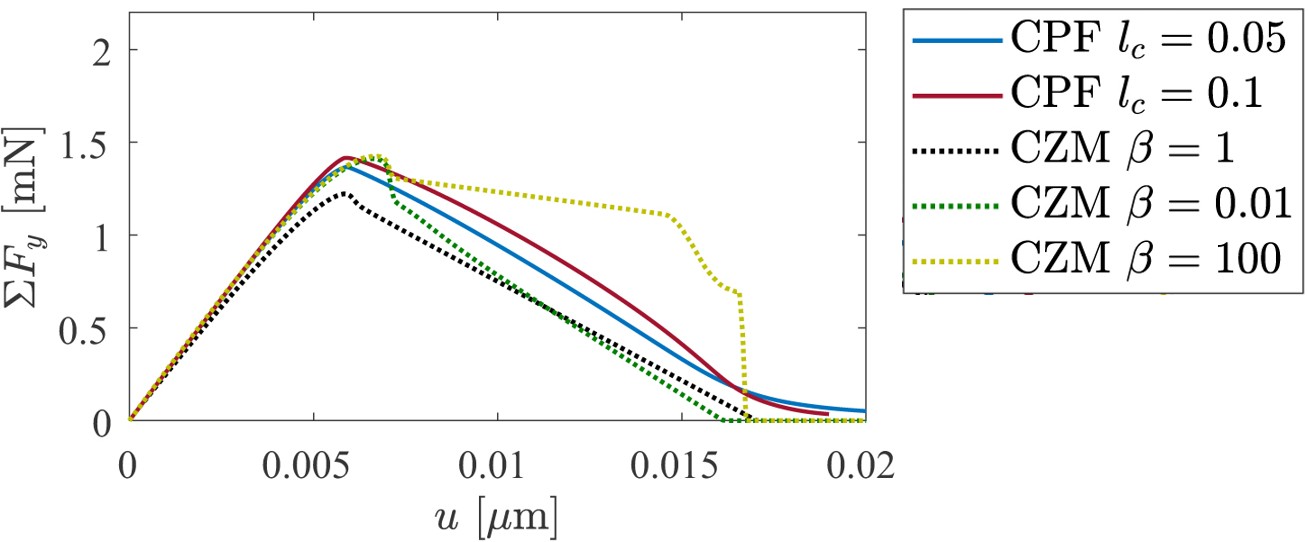}
    \caption{Comparison between CZ with different $\beta$ and CPF. }
    \label{fig:anSCPF_S_RF}
\end{figure}

In the next step, we studied the same anisotropic cracking utilizing the CZ model. Here, we take advantage of the PF fracture results to determine in which direction the crack propagates ($\theta_c \approx 20^{\circ}$). The very same plane is enriched with CZ elements introduced previously. Moreover, we also studied the influence of fracture mode-mixity. In other words, the parameter $\beta$ in the CZ formulation (see Eq.~\ref{eq:cz_biliear}) is varied. Choosing a relatively small value, i.e. $\beta=0.01$, indicates a weak contribution from the shear direction upon shear opening. Furthermore, choosing $\beta=1.0$ means isotropic behavior for the CZ formulation. Finally, by setting $\beta=100$, the contribution of the shear traction is much more pronounced. 
The results of the comparison between the CPF and CZ model with different mode-mixity parameters are summarized in Fig.~\ref{fig:anSCPF_S_RF}. First, we observe that the results of the CPF model are almost length-scale insensitive even in the case of anisotropic fracture. Second, the results of the CZ model match very well with the CPF model only if $\beta=1.0$ or $\beta=0.01$. In other words, when the contribution of the shear traction is much more due to the fracture mode-mixity (i.e. $\beta=100$), the post-fracture results of the CZ model deviate from those of CPF. The later point opens up the necessity of taking into account the mode-mixity into the PF damage formulation. By doing, one can perhaps can think about utilizing more damage variables for each fracture mode \cite{FEI2021113655}. See also \cite{SHANTHRAJ201719, BRYANT2018561}.

\textbf{Remark 7.} Despite being consistent with the CZ models, the cohesive phase-field formulation still lacks one of the main features of CZ models which is the mode-dependent nature of the fracture. The latter point should be studied in future developments. One idea would be to consider multiple damage variables to represent different fracture modes \cite{FEI2021113655}. 


\subsection{Anisotropic crack propagation utilizing a direction-dependent fracture energy} 
In this section, we look at anisotropic cracking in specimens described in Table~\ref{tab:par_geo}, this time by utilizing the formulation based on arbitrary anisotropy (i.e. Eqs.~\ref{psiarbit}, \ref{aransio} and \ref{rda}). The model parameters regarding anisotropic fracture ($\alpha_m$ and $\theta_m$) are reported in Table~\ref{tab:par_mat}.
Furthermore, in the current simulations we propose $||\nabla d||_c=0.04~l_c$. It is checked that this parameter is small enough so the numerical solver converges and the obtained results remain unchanged with respect to this  (see also studies in \cite{REZAEI2021a}).

\begin{figure}[H]
    \centering
    \includegraphics[width=1.0\linewidth]{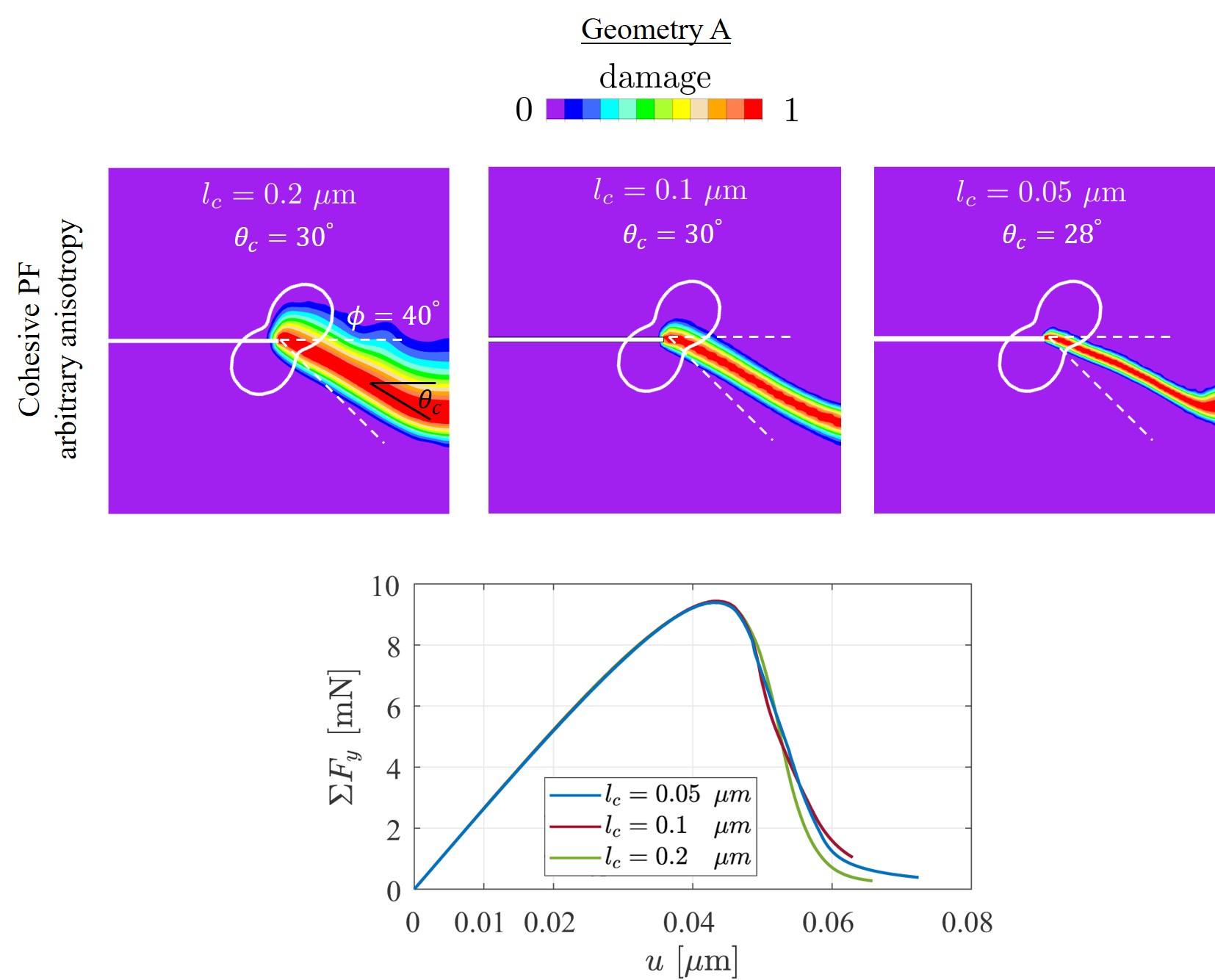}
    \caption{Studies on anisotropic crack propagation using the introduced anisotropic CPF damage formulation. }
    \label{fig:anACPF_L}
\end{figure}

The results of the obtained crack path are plotted in Fig.~\ref{fig:anACPF_L} for different values of the length scale parameter. Similar to the previous study, the crack path angle converges to a certain value $\theta_c \approx 30^{\circ}$ for all the cases. The reaction forces are shown in the lower part of Fig.~\ref{fig:anACPF_L}. Interestingly enough, utilizing the CPF model, the obtained reaction forces are very similar which shows the almost insensitive response of the formulation with respect to the length scale parameter. 

These results are obtained based on an arbitrary function of the fracture energy as well as the strength of the material. In other words, with one single damage variable one can take into account any complicated fracture energy distribution. The latter point is an efficient way to simulate anisotropic cracking in many available materials (see also \cite{REZAEI2021a}).


\textbf{Remark 8.} Despite the nearly length scale insensitive results we cannot simply choose $l_c$ as large as we want. According to \cite{lorentz2017,wu2018}, for having numerical stability, $a_1 \geq \dfrac{3}{2}$ should be fulfilled, which means $l_c \leq 0.85~l_{ch}$. Here, $l_{ch}=\dfrac{EG_c}{\sigma_0^2}$ is the characteristic length scale of the problem. Therefore, there is an upper limit for $l_c$.

\subsection{Three-point bending test with anisotropic properties}
\begin{figure}[H]
    \centering
    \includegraphics[width=1.0\linewidth]{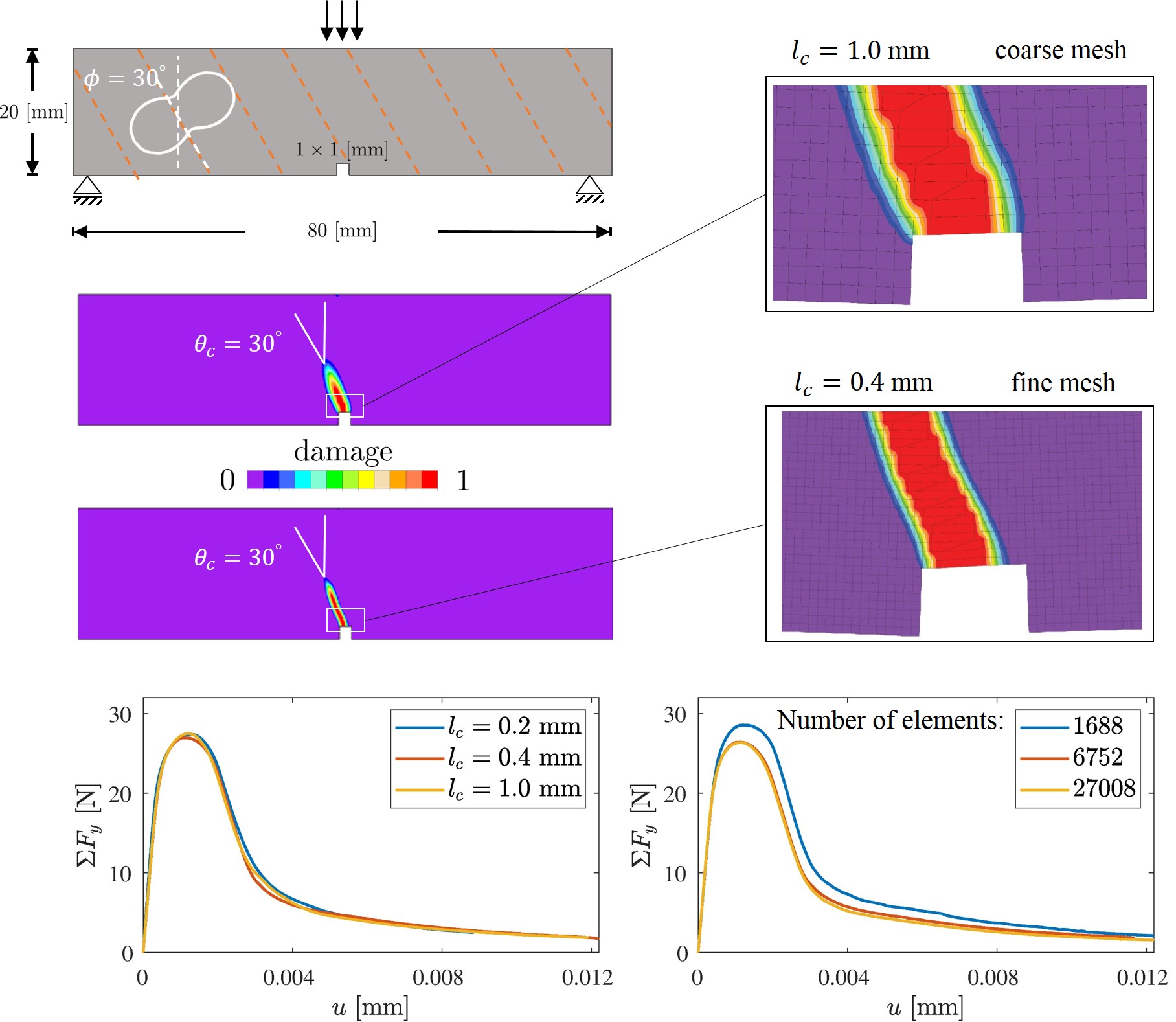}
    \caption{Studies on the three-point-bending test. For two different length scale parameters $l_c$, the results regarding the obtained crack paths as well as overall reaction forces are compared.}
    \label{fig:bending}
\end{figure}

The geometry and material parameters for this test are shown in Fig.~\ref{fig:bending}. The material direction which can be interpreted as fibers direction or a layered material is represented by the angle $\phi=30^{\circ}$. For more realistic calculations, the anisotropic elastic properties are also considered for this example by having the grain orientation depicted in Fig.~\ref{fig:bending}. The anisotropic elastic and fracture properties are reported in Table~\ref{tab:par_matbending}, see also \cite{Nejati2020}. This problem is solved utilizing the introduced anisotropic CPF model with arbitrary function for the fracture energy distribution (similar to Section 4.3).
\begin{table}[H]
	\centering
	\begin{tabular}{ l l l } \hline
	\multirow{1}{*}{}         & Unit    & Value    \\ \hline \hline 
	Elastic constant $C_{11}$ & [MPa] & $142350$  \\
	Elastic constant $C_{12}$ & [MPa] & $188782$\\
    Elastic constant $C_{16}$ & [MPa] & $115880$\\
    Elastic constant $C_{26}$ & [MPa] & $192680$\\
    Elastic constant $C_{22}$ & [MPa] & $321110$\\
    Elastic constant $C_{66}$ & [MPa] & $126370$\\
	Fracture energy $G_{c,0}$  & [$\dfrac{\text{J}}{\text{m}^2}$] $\equiv$ [MPa.$ \text{mm}$]$10^3$ & $54$\\
	Ultimate strength $\sigma_0$ $=\sigma_{0,1}$  & [MPa] & $10$\\
	Damage internal length $l_c$    & [mm]  & $0.4-1.0$   \\ 
	Frequency number $m$            & [-]   & $1$  \\ 
	Fracture energy parameter $\alpha_m$     & [-] & $3.5$ \\ 
	Fracture energy parameter $\theta'_m$    & [-] & $60^\circ$ \\
    Damage parameter $||\nabla d||_c$  & [-] & $0.2$ \\
    Material strength parameter $p_m$ & [-] & $0.1$    \\
    \hline
	\end{tabular}
	\caption{Parameters for the anisotropic PF damage formulation and anisotropic material.}
	\label{tab:par_matbending}
\end{table}
A displacement on the top edge is applied and the reaction forces are measured accordingly. As expected, due to the anisotropic properties the crack direction runs along the angle $\phi$. For the chosen material properties, the obtained crack path is very close to this preferential crack angle, i.e. $\theta_c \simeq 30^{\circ}$. We also study the influence of the length scale parameter $l_c$. For the chosen values, not only the final crack paths but also the overall measured reaction forces are in very good agreement (see the lower part of Fig.~\ref{fig:bending}). To ensure the accuracy of the obtained results, a mesh convergence study is performed for the case with $l_c=2$~mm. See also similar studies in the context of rock mechanics \cite{Vowinckel2021} and also when it comes to fiber composite materials \cite{schreiber2021,ZHANG2019105008} utilizing standard PF damage models.
It is worth mentioning that, depending on the chosen length scale parameter, the element size for the finite element calculation can be changed to reduce computational time. In the current studies the computational cost of the simulation with $l_c=1.0$ mm is almost half the case with $l_c=0.4$ mm. The latter point shows another advantage of the length scale insensitive formulation and its flexibility for choosing the length scale parameter. Nevertheless, one has to keep in mind the restrictions described in Remark 8. Depending on the size of the specimen, this parameter should not be chosen too large, otherwise the crack path might get too diffused which may not be physically accurate.

\subsection{Anisotropic cracking in crystalline materials with diffuse interphase}
\begin{figure}[H]
    \centering
    \includegraphics[width=1.0\linewidth]{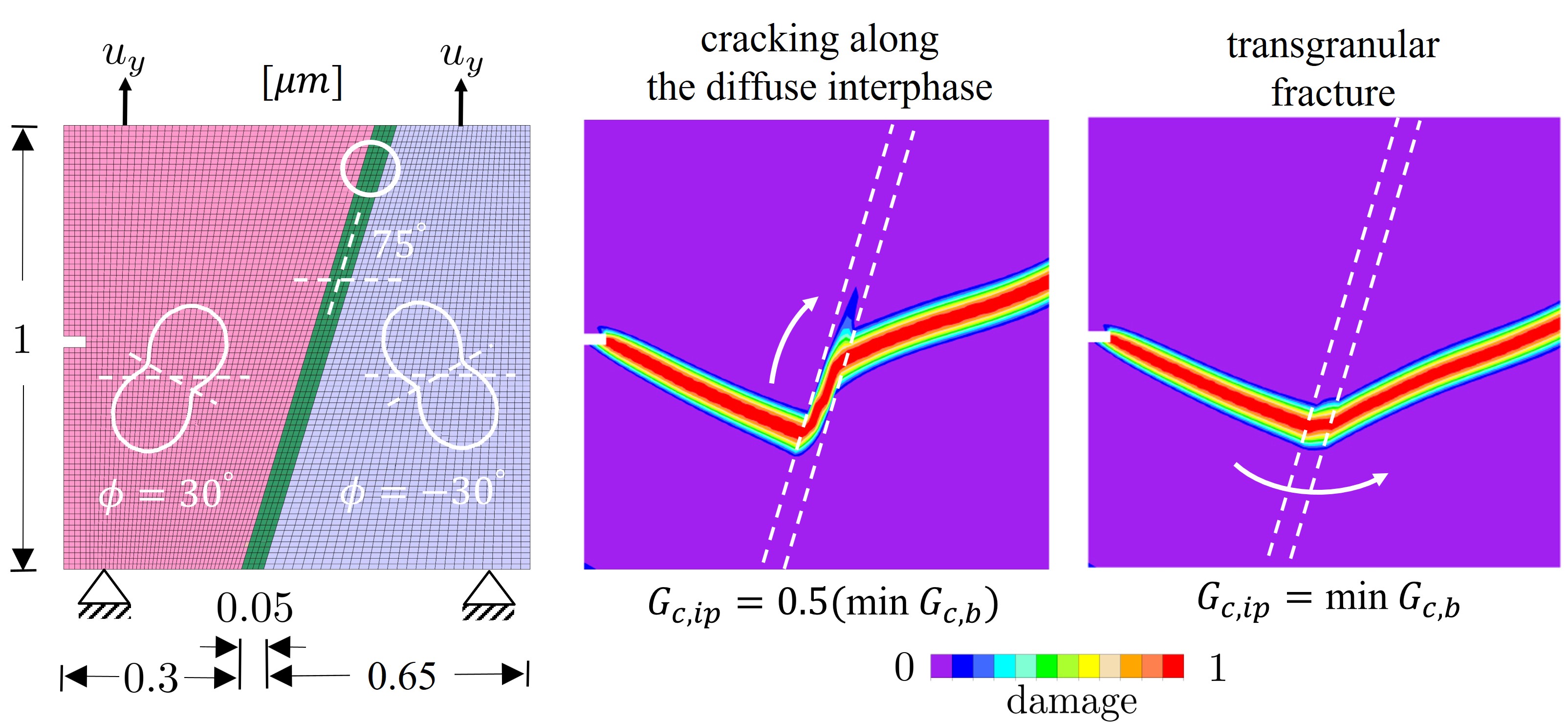}
    \caption{Studies on anisotropic cohesive fracture within a bi-crystalline system. The grain boundary is treated as a diffuse zone. The introduced anisotropic cohesive phase filed model can handle cracking within the bulk and interphase.}
    \label{fig:bicr}
\end{figure}

To show the potential of the cohesive phase-field approach, we discuss the cracking in a simple bi-crystalline system according to Fig.~\ref{fig:bicr}. Here, the grain boundary is represented by a diffuse zone in green color. The two neighboring grain each has specific orientation as shown in the figure. For the diffuse interphase, an anisotropic distribution for the fracture energy is considered which its orientation is exactly set according to the grain boundary angle (i.e. $75^{\circ}$). All the anisotropic cohesive phase field formulations are based on structural tensor according to Eqs.~\ref{anisostpsic}, \ref{deltadsttensor} and \ref{rds}. Other material properties such as elastic modulus and fracture properties are according to Table~\ref{tab:bicryst_mat}.
\begin{table}[H]
	\centering
	\begin{tabular}{ l l l } \hline
	\multirow{1}{*}{}         & Unit    & Value    \\ \hline \hline 
	Lamé's Constants ($\lambda,~\mu$)   & [GPa] & $(132.6, 163.4)$  \\
	Bulk fracture energy $G_{c,b}$   &  [GPa.$\mu \text{m}$]$10^3$ & $30$\\
	Interphase fracture energy $G_{c,ip}$   &  [GPa.$\mu \text{m}$]$10^3$ & $30$,~$15$\\
	Bulk ultimate strength $\sigma_{0,b}$ $=\sigma_{0,1}$  & [GPa] & $4$\\
	Interphase ultimate strength $\sigma_{0,ip}$ $=\sigma_{0,1}$ & [GPa] & $4$,~$2$\\
	Damage internal length $l_c$    & [$\mu$m]  & $0.025$   \\ 
    Structural parameter $\alpha$    & [-] & $12.0$ \\
    Structural parameter $\phi$    & [-] & $-30^\circ,~30^\circ$ \\
    \hline
	\end{tabular}
	\caption{Parameters for the anisotropic PF damage formulation.}
	\label{tab:bicryst_mat}
\end{table}
Note that in this example, there is no need for the insertion of additional cohesive zone elements. In other words, the cohesive phase-field approach on its owns includes the same properties. By applying displacement in the vertical direction on the top edge, crack propagation in this system is studied. For similar studies readers are encouraged to see \cite{Nguyen2017, PAGGI2017145, REZAEI2021a, LOTFOLAHPOUR2021110642}. 

For a better comparison, the fracture energy value for the interphase $G_{c,ip}$ is varied against the one for the bulk part $G_{c,b}$. In the middle part of Fig.~\ref{fig:bicr}, the results for the weaker grain boundary are shown, where the crack tends to propagate along the interphase and then goes to the other grain. On the other hand, by increasing the interphase fracture energy, as shown in the right-hand side of Fig.~\ref{fig:bicr}, the transgranular fracture is observed. 


\section{Conclusion and future work}
In this contribution, we try to address anisotropic cohesive fracture using the phase-field damage model. In other words, direction-dependent damage initiation and propagation within an arbitrary anisotropic solid are under focus.

It is well established that standard PF damage models provide a consistent formulation that can predict crack initiation and propagation. By dissipating the fracture energy within a diffuse zone controlled by the length scale parameter, these models solve the problem of mesh sensitivity during damage progression. The length scale parameter, on the other hand, has a significant influence on the global response of the model. This parameter is shown to be related to the maximum strength of the material and, therefore, can control damage nucleation. We discuss that the latter point is not desirable for all applications, especially when the size of the specimen is not large enough compared to the internal length scale parameter. Furthermore, for high-strength materials, the mesh has to be extremely refined which increases the computational costs significantly.

Firstly, the sensitivity of the system's global response with respect to the length scale parameter is shown for standard PF damage formulation. Secondly, an insensitive formulation \cite{LORENTZ201120, wu2018, GEELEN2019} is adopted and then extended for the anisotropic case. In particular, we focus on utilizing the direction-dependent fracture energy formulation \cite{REZAEI2021a} and second-order structural tensors \cite{Teichtmeister17}. Considering the numerical implementation, a linearization procedure and details of utilized algorithms are discussed as well. The crack initiation and propagation in a single notched specimen with two different geometries as well as a simple three-point bending test are studied. It is shown that the formulation can produce almost insensitive results with respect to the length scale parameter for both isotropic and anisotropic cases. We also compared the numerical results against results obtained by studying cracking using the standard cohesive-zone model. It is shown that the framework can reproduce the results from the CZ formulation, especially when there is no severe difference between different opening modes behavior.

We conclude that the cohesive phase-field formulation has two main advantages: we include more (clear) physics into account by introducing the strength and fracture energy as input parameters. In other words, the length scale parameter can be treated as a numerical parameter which should be small enough, depending on the application and the boundary value problem. The latter point is extremely helpful in multiphysics problems where the fracture properties are under the influence of other fields \cite{MANDAL2021102840}. Furthermore, one can relatively increase the mesh size. We show that the latter point reduces the computational time significantly without any severe change in the predicted crack path or overall obtained load-displacement curves.

The developed model can be applied in efficient numerical modeling of fracture at smaller scales. For example, see studies by \cite{REZAEI2018192, Khaledi2018b} on micro-coating layers where the thickness of the coating system is about a few micrometers which are in order of the obtained internal length. Therefore, it not so easy to simulate the problem with standard PF damage models. 

Further comparisons with similar models can be very interesting to complete our understanding of the anisotropic nonlocal fracture in solids. As an example, the PF damage formulation benefits from regular mesh generation, while cohesive zone models suffer from predefined crack path zones and a specific mesh algorithm. Also, comparisons with other methodologies such as XFEM and Peridynamics would certainly be interesting.

Apart from the advantages of the current anisotropic CPF formulation, there are some open issues and possibilities for further improvements. We showed that CPF models still lack to capture mode-dependent fracture properties.
A crucial enhancement for the formulation could be made to consider different modes of opening. As a possible remedy, one could introduce different damage variables for each opening mode. Utilizing multiple damage variables would cause a degrading of the elasticity matrix components with different damage values. Meanwhile, multiple damage variables could be beneficial and enable the model to capture different stresses and anisotropic responses \cite{REESE2021, FEI2021113655}. Another idea for further developments would be to degrade the fracture toughness value to represent fatigue behavior \cite{YIN2020113068}. Finally, extension to large deformation and including plasticity into the damage formulation is of great interest \cite{BREPOLS2020102635}.\\ \\
\textbf{Acknowledgements}:~Financial support of Subproject A6 of the Transregional Collaborative Research Center SFB/TRR 87 and Subproject A01 of the Transregional Collaborative Research Center SFB-TRR 280 by the German Research Foundation (DFG) is gratefully acknowledged.

\section{Appendix A: Analytical solution for $1$-D damage sub problem}
In this appendix, a closed-form solution for $1$-D damage PDE is presented. By which the difference between standard and cohesive PF models, can be interpreted. Readers are also encourage to see \cite{wu2017, GEELEN2019}. To simplify the equations following function is introduced as:
\begin{equation}
  \label{gdf}
g(d)= \dfrac{1}{{f_D}} -1 \Rightarrow g^{\prime}(d) = \dfrac{-{f^{\prime}_D}}{{f_D}^2}.
  \end{equation}
Which reads:
\begin{equation}
  \label{sigeps}
\epsilon(d)=\dfrac{\sigma}{E_0}{f_D}^{-1}=\dfrac{\sigma}{E_0}(g(d)+1).
  \end{equation}
With having damage PDE in one hand and the $1$-D elastic energy as $\psi_{e,1D}=\dfrac{1}{2}E\epsilon^2$, the PDE of damage can be rewritten as:
\begin{equation}
  \label{g1d}
\dfrac{\sigma^2 g^{\prime}(d)}{2E_0}-\dfrac{G_c}{c_0l_c}\left(\omega^{\prime}(d) -2 l_c^2 d_{,xx} \right)=0.
  \end{equation}
Assuming uniform damage, the latter term can be neglected.
\begin{equation}
  \label{g2d}
 \dfrac{\sigma^2 g^{\prime}(d)}{2E_0}-\dfrac{G_c}{c_0l_c} \omega^{\prime}(d) =0
\end{equation}
As a result, the following equations are obtained for strain and stress at the onset of crack initiation ($d=0$):
\begin{equation}
\label{streps1d}
\begin{cases}
\sigma=\sqrt{\dfrac{2E_0G_c}{c_0l_c}~\dfrac{\omega^{\prime}(0)}{g^{\prime}(0)}}\\
\epsilon=\dfrac{1}{E_0}\sqrt{\dfrac{-2E_0G_c}{c_0l_c}~\dfrac{\omega^{\prime}(0)}{{f^{\prime}_D}_c}}.
\end{cases}
\end{equation}
The above expressions are obtained, by using  L’Hôpital’s rule since the limit is indeterminate ($ \lim_{d\to0} \dfrac{\omega(d)}{g(d)}=\frac{0}{0})$.

A linear term in the crack topology function yields an initial elastic stage before damage initiation, and the maximum stress is achieved, when $d=0$. On the contrary and in the standard phase-field approach, the damage initiates from infinitesimal tensile strain, and stress reaches its maximum value when $d=0.25$. Recalling Eq.~\ref{streps1d}, $a_1$ computed as
\begin{equation}
\label{t0alphaphi}
\sigma=\sqrt{\dfrac{2E_0G_c}{c_0l_c}~\dfrac{2}{a_1}} ~~~\Rightarrow~~~a_1=\dfrac{2EG_c}{\sigma_0^2c_0l_c},
\end{equation}  
which grantees the value of maximum stress to be $\sigma_0$ independently of the internal length scale $l_c$. 
Considering the TSL (depicted in Fig. \ref{fig:LvsS}), $\lim_{[[u]]\to\lambda_f}{\sigma([[u]]})=0$ is accomplished with having the final crack opening as:
\begin{equation}
\label{Wc}
W_u= \dfrac{2 \pi G_{c}}{\sigma_0 c_0} \sqrt{2(1+a_2)}.
\end{equation} 
Having the $c_0=\pi$ and $\lambda_f=\dfrac{2 G_{c,0}}{\sigma_0}$ in hand, for fulfilling $\lambda_f=W_u$, we have 
\begin{equation}
\label{a2}
a_2=-0.5.
\end{equation} 


\section{Appendix B: Regularised crack density function} 
In this appendix we provide and review some information regarding a general form of the crack density function introduced in Eqs.~\ref{stcrackdensity} and \ref{arcrackdensity}
\begin{equation}  
\label{crckdnsalfa}
\gamma (d,\nabla d)~=~\dfrac{1}{c_0}\left(\dfrac{\omega(d)}{l_c} + l_c \nabla d \cdot \nabla d \right)
\end{equation}
According to Euler-Lagrange principle, the governing equation for phase-field damage is obtained as
\begin{equation}  
\label{condiomega}
\begin{cases}
\dfrac{\text{d}\omega(d)}{\text{d}d}-2l_c^2~ \Delta d =0~~~ \text{in}~~\Omega\\
\nabla d \cdot n =0~~~~~~~~~~~~~~~\text{on}~~\partial \Omega
\end{cases}
\end{equation}
By multiplying the above equation by~$d^{\prime}$~ and integrating along the normal direction to the crack direction one obtains: 
\begin{equation}  
\label{regeq}
\begin{cases}
\omega(d)-l_c^2 {|\nabla_n|}^2=0 \Rightarrow~\gamma=\dfrac{2}{c_0l_c}\omega(d)\\
{|\nabla_n|}:=\dfrac{\partial d}{\partial |x_n|}=\dfrac{1}{l_c}\sqrt{\omega(d)}
\end{cases}
\end{equation}
Here, we defined $x_n:= \left(\bm{x}-\bm{x}_c\right) \cdot \bm{n}_c$, where $x_n$ is scalar product of the vector which obtained as the distance of point $x$ from its closest point at the surface of crack $x_c$. The normal vector to the crack surface is denoted by $\bm{n}_c$. \color{black}
Considering~$dV=2|dx_n|\cdot A_s$ and Eq.\ref{arcrackdensity} reads:
\begin{equation}  
\label{adgamamd}
\Gamma_{c} = \int_B \gamma dV = \dfrac{4}{c_0}\int_0^d \omega(d) ~\dfrac{1}{l_c} ~d|x_n|\cdot A_s
\end{equation}
where~$A_s$~ is the surface of crack. Finally, it follows as:
\begin{equation}  
\label{c0formula}
\Gamma_{c}=A_s \Rightarrow~ c_0=4\int_0^d \omega(d)~\dfrac{1}{l_c} ~d|x_n|=4\int_0^1\sqrt{\omega(\beta)}d\beta
\end{equation}
\color{black}
Different groups of function can be chosen for $\omega(d)$, nevertheless they should fulfill the following conditions: $\omega(0)=0,~
\omega(1)=1,~
\omega^{\prime}\geq 0$.
Some choices for crack geometric function in PF damage models are \cite{Miehe2010a,Pham2011,Bourdin2014,wu2017}:
\begin{align}
\label{geomd2}
\begin{cases}
\omega(d)=d^2\\
d_u(x)= \exp\left({\dfrac{-|x|}{l_c}}\right)\\
D_u=+\infty \\
c_0=2
\end{cases},~~~~~
\begin{cases}
\omega(d)=d\\
d_u(x)= \left(1-\dfrac{-|x|}{2l_c}\right)^2\\
D_u=2l_c \\
c_0=\dfrac{8}{3}
\end{cases},~~~~~
\begin{cases}
\omega(d)=2d-d^2\\
d_u(x)= 1-\sin\left(\dfrac{|x|}{l_c}\right)\\
D_u=\dfrac{\pi}{2}l_c \\
c_0=\pi
\end{cases}
\end{align}
Note that by having the linear term in the crack topology function, one can introduce the threshold for damage. In other words, for the first choice ($\omega(d)=d^2$), damage zone expands towards infinity.
Where, $D_u$ denotes to the damage half bandwidth.

\section{Appendix C: Derivation of anisotropic crack energy using structural tensor}
The expression in Eq.~\ref{anisostpsic} for the crack energy using the structural tensor $\bm{A}$ is derived in this appendix. Note that $c(\bullet)$ and $s(\bullet)$ denote the functions $\cos(\bullet)$ and $\sin(\bullet)$, respectively.
Recalling Eq.~\ref{regularizedcrackenergy} and the definition of $\bm{A}$ in Eq.~\ref{Atensor} we have the following equation for the fracture energy. 
\begin{align}  
\label{psicAa}
\psi_{c,a} &= \dfrac{G_{c,0}}{c_0l_c} ~\left(\omega(d)~+~l_c^2~\nabla d^T \, \bm{A} \, \nabla d  \right) \\
&=\dfrac{G_{c,0}}{c_0l_c} ~\left(\omega(d)~+~\alpha l_c^2~\nabla d^T \, \begin{bmatrix}
c^2(\phi) & c(\phi)s(\phi)\\
c(\phi)s(\phi) & s^2(\phi)
\end{bmatrix} ||\nabla d||^2 \, \nabla d  ~+~l_c^2~\nabla d^T~\nabla d \right).
\end{align}
Considering Eq.~\ref{theta}, the direction of $\nabla d$ is denoted by the angle $\beta=\text{atan}\left( \dfrac{\nabla d \cdot e_2}{\nabla d \cdot e_1} \right)=\theta +\pi/2$.
Therefore, we have $\nabla d^T = \begin{bmatrix} c(\beta)~~  s(\beta) \end{bmatrix}$. One can further simplify the above expression as
\begin{align}
\psi_{c,a}  &=\dfrac{G_{c,0}}{c_0l_c} \left(\omega(d)+\alpha l_c^2~\begin{bmatrix} c(\beta)& s(\beta)  \end{bmatrix} \begin{bmatrix} c^2(\phi)\,c(\beta)\,+\,c(\phi)\,c(\beta)\,s(\beta) \\ c(\phi)\,s(\phi)\,c(\beta)\,+s^2(\phi)\,s(\beta) \end{bmatrix}||\nabla d||^2 \,+\,l_c^2~\nabla d^T~\nabla d \right), \\
&=\dfrac{G_{c,0}}{c_0l_c} ~\left(\omega(d)~+~\alpha l_c^2~\left(c(\phi)c(\beta)+s(\phi)s(\beta)  \right)^2 \,+\,l_c^2 \nabla d^T \nabla d \right), \\
&=\dfrac{G_{c,0}}{c_0l_c} ~\left(\omega(d)~+~\alpha l_c^2~\cos^2(\beta-\phi)||\nabla d||^2 \,+\,l_c^2 \nabla d^T \nabla d \right).
\end{align}
By reconsidering $\beta\,=\,\theta\,+\,\pi/2$, we have:
\begin{align}
\psi_{c,a} &=\dfrac{G_{c,0}}{c_0l_c} ~\left(\omega(d)~+~\left(1\,+\,\alpha l_c^2~\sin^2(\theta-\phi)\right)||\nabla d||^2\right).
\end{align}


\newpage
\bibliographystyle{elsarticle-num}
\bibliography{bib}

\end{document}